\documentclass[12pt]{article}
\usepackage{amsmath,amssymb,amsfonts}
\usepackage{graphicx}

\newcommand{\beq}{\begin{equation}} 
\newcommand{\eeq}{\end{equation}} 
\newcommand{\bea}{\begin{eqnarray}}  
\newcommand{\eea}{\end{eqnarray}}  

\newcommand{\refeq}[1]{\mbox{(\ref{#1})}}

\begin{document}


\begin{flushright}
UG-FT-224/08 \\
CAFPE-94/08 \\

\today
\end{flushright}
\vspace*{5mm}
\begin{center}

\renewcommand{\thefootnote}{\fnsymbol{footnote}}

{\Large {\bf Effects of new leptons in Electroweak Precision Data
}} \\
\vspace*{1cm}
{\bf F.\ del Aguila}\footnote{E-mail: faguila@ugr.es},
{\bf J.\ de Blas}\footnote{E-mail: deblasm@ugr.es}
and
{\bf M.\ P\'erez-Victoria}\footnote{E-mail: mpv@ugr.es}

\vspace{0.5cm}

Departamento de F\'{\i}sica Te\'orica y del Cosmos and CAFPE,\\
Universidad de Granada, E-18071 Granada, Spain

\end{center}
\vspace{.5cm}
\begin{abstract}
 
\noindent We obtain limits on generic vector-like leptons at the
TeV scale from electroweak precision tests. These limits are
complementary to the ones obtained from lepton flavour violating
processes. In general, the quality of the global electroweak fit is
comparable to the one for the Standard Model. In the case of an extra
neutrino singlet mixing with the muon or electron, the global fit
allows for a relatively large 
Higgs mass ($M_H\lesssim 260~\mathrm{GeV}$ at 90\,\% C.L.), thus
relaxing the tension between the direct LEP limit and the Standard
Model fit.

\end{abstract}

\renewcommand{\thefootnote}{\arabic{footnote}}
\setcounter{footnote}{0}

\section{Introduction}

In this paper we analize the effects of new vector-like leptons in
electroweak precision data (EWPD). We derive the corresponding
effective Lagrangian up to dimension 6, and use it to study these
effects from two 
related points of view. On the one hand, we obtain limits on the
couplings and masses of the new particles. On the other, we observe
that the new leptons can mildly improve some features of the Standard
Model (SM) fit and have consequences on the preferred values of the
Higgs mass $M_H$. The largest limit on $M_H$ arises for new
leptons transforming as neutrino singlets under the SM. If they have
Majorana masses, they may act as see-saw messengers and the
restrictions from the limits on light neutrino masses must be taken
into account. 

Let us first review shortly the situation in the SM. As it is well known,
EWPD are consistent with the SM to a remarkable precision, sensitive
to the details of radiative  
corrections \cite{Yao:2006px}. Despite this general success, a few
experimental results are difficult to accommodate within the SM
picture. The discrepancies could have experimental origin, but it is
nevertheless interesting to study them in some detail, to see if they
follow some pattern and could give us some hint of new physics. 
The main problem at the Z pole is
with the value of $\sin^2 \theta_\mathrm{eff}^\mathrm{lept}$, which is
distinctively higher 
when derived from hadronic asymmetries than when derived from the
leptonic ones.  The statistical probability that the set of asymmetry data be
consistent with the SM hypothesis is only 3.7 \%~\cite{Grunewald:2007xt}. This
low probability is driven by the two most precise determinations of $\sin^2
\theta_\mathrm{eff}^\mathrm{lept}$, obtained from the leptonic
asymmetry parameter $A_l$ by SLD and of the bottom 
forward-backward asymmetry $A_\mathrm{FB}^{0,b}$ at LEP,
respectively. These measurements differ by 3.2 
standard deviations ($\sigma$). On the other hand, the SM prediction depends
through quantum corrections on the
unknown value of the Higgs boson mass $M_H$, and agrees with leptonic
(hadronic) data for a light (heavy) Higgs. The current global fit
in~\cite{EWWG} to Z-pole observables plus the masses of the top
quark $m_t$ and W boson $M_W$, and the W width $\Gamma_W$, prefers a
light Higgs: $M_H=87^{+36}_{-27} \,
\mathrm{GeV}$. Note that this conclusion does not seem
compelling, as it arises from the 
combination of contradictory measurements: 
$M_W$ and the leptonic asymmetries at the Z pole point to a
very light Higgs, whereas the hadronic asymmetries prefer a heavy
Higgs~\cite{Chanowitz:2002cd}. 
At any rate, this best-fit value gives a prediction for
$A_\mathrm{FB}^{0,b}$ that is $2.9$ $\sigma$ above its experimental
value, while 
the leptonic asymmetries differ by less than $1.6$
$\sigma$~\cite{EWWG}. For this reason, it is common to speak of a
$A_\mathrm{FB}^{0,b}$ 
anomaly, and implicitly consider that the leptonic data are in good
agreement with the SM.
One should not forget, however, that LEP~2 has put a sharp
limit on the mass of the SM Higgs boson: $M_H\geq 114.4\,\mathrm{GeV}$ (95\%
C.L.)~\cite{Barate:2003sz}. With this constraint, the best SM fit
is realized for the lowest allowed $M_H$. Then, we find that the pulls in
$A_\mathrm{FB}^{0,b}$ and $A_e$(SLD) are, respectively, $2.6$ and $2.0$.  

We use a data set including low-$Q^2$ measurements. In
Tables~\ref{Exp-SM} and~\ref{Exp-SM-LU} in the appendix,
we collect the experimental values of different  
(pseudo) observables at different energies, together with the
corresponding predictions and 
pulls in the SM for $M_H=114.4\,\mathrm{GeV}$. We use the new
(preliminary) CDF-D\O\ value for the top mass, $m_t=172.6 \pm 1.4\, 
\mathrm{GeV}$~\cite{Tevatronmt}. We find
$\chi^2/\mathrm{d.o.f.}=43.9/30$, which 
corresponds to a probability of 4.8\,\% only. More details on   
this fit are given below. For the moment, let us just point out
the main discrepancies between experiment and the SM, beyond the ones
stressed above. First, there is a $2.8\ \sigma$ discrepancy, coming
from the NuTeV experiment, in the
effective coupling $g_L^2$ that enters neutrino-nucleon
scattering. Unexpectedly 
large isospin violations~\cite{LIV} or a significant quark-antiquark
asymmetry in the strange sea quarks~\cite{SAS} could account for part of the
deviation, but it seems difficult to explain the whole effect with
standard physics only. Second, the pulls of $M_W$ and the hadronic
cross section at the Z pole $\sigma_H^0$ are at 
$1.3\ \sigma$ and $1.7\ \sigma$, respectively. And finally, the data
show departures from lepton  
universality in both $Z$ and $W$ decays\footnote{There is also a large
discrepancy in the anomalous magnetic moment of the muon
$g_\mu-2$~\cite{magneticmoment}, but 
we do not include this observable in our fit because the contributions
of the extra leptons to it are smaller than the experimental and
theoretical errors. Nevertheless, at the end of
Section~\ref{section_globalfit} we 
comment on some subtle implications of $g_\mu-2$ through the value
of the parameter $\Delta \alpha_\mathrm{had}^{(5)}(M_Z^2)$.}. There have 
been several attempts to explain some of these deviations (mainly
the $A_\mathrm{FB}^{0,b}$ or NuTeV anomalies) with new physics, see
for example~\cite{Choudhury:2001hs,Loinaz}. In this
regard, it is important to be careful that the new physics that
corrects a particular observable does not spoil the goodness of the
global fit, and also to have into account the direct LEP~2 lower bound
on $M_H$.

Here we study the impact that new
fermionic $SU(3)_c$ singlets (leptons) have on EPWD. 
Since these hypothetical particles modify lepton
observables, it looks plausible, a priori, that they may improve the
electroweak fit and/or change the prediction for the Higgs
mass. We consider all possible new colour-neutral vector-like fermions that,
after electroweak symmetry breaking, mix with the SM neutrinos or
charged leptons, and hence contribute to precision observables. These
new leptons are
predicted in many theories beyond the SM, including Grand Unified
Theories (GUT)~\cite{GUTs}, models in extra dimensions~\cite{XDs} and
Little Higgs models~\cite{LH}. As they are relatively heavy,
an effective Lagrangian approach should be a good
approximation. In fact, we will integrate out the new leptons keeping
only the operators up to dimension~6. The use of an effective formalism
to fit EWPD also allows for a common treatment for any kind
of new physics~\cite{skiba}. We leave a more general analysis
for future work~\cite{In preparation}.

We find that the quality of the global fit (including high- and low-$Q^2$
data) hardly improves when the new leptons are
included. The case of neutrino singlets has the interesting feature of
raising the preferred Higgs mass to a confortable region above the
direct LEP limit. Due to the values of $M_W$ 
measured at LEP~2 and Tevatron, however, the Higgs cannot be very
heavy. For other kinds of new leptons, the SM prediction for $M_H$ is
mostly unchanged. 

From the global fits, we extract limits on the
mixings of the different possible new leptons with the SM
ones. (The limits that had been obtained before for some of these heavy
leptons in~\cite{Langacker:1988ur,Nardi:1994iv,Bergmann:1998rg} are
improved.)  The upper
bounds on the allowed mixings range from 0.01 to 0.08 at 90 $\%$ C.L.,
depending on the quantum numbers of the new lepton and the family of
the SM lepton it mixes with. 
If the new leptons are 
weakly coupled, the largest allowed mixings require that their masses be not
far from the TeV scale.

It is important to note that new leptons with significant mixings
are generically ruled out when they mediate Flavour Changing
Neutral Currents (FCNC)~\cite{Tommasini:1995ii,Abada:2007ux,Raidal:2008jk}, 
generate masses for the SM neutrinos~\cite{delAguila:2007ap} 
or contribute to neutrinoless double $\beta$
decay~\cite{Mohapatra:1998ye}.  
To avoid these constraints, we must assume that each new lepton mixes
mostly with just one family, and that their contributions to the light
Majorana masses  
and neutrinoless double $\beta$ decay, when
allowed, are very suppressed~\cite{Ingelman:1993ve}. 
This scenario with new Majorana particles at the
TeV scale that have sizeable mixings with the SM leptons can be made
natural with the help of extra symmetries. 
In general, these include
lepton number (LN) conservation~\cite{Kersten:2007vk} and must be
very slightly broken, if at all. At any rate, we
find that new leptons with the quantum numbers of see-saw
messengers of type I~\cite{seesaw} and III~\cite{seesawIII} and
sizable mixings can be consistent with EWPD. The neutrino singlets
are also relevant to models of resonant leptogenesis~\cite{leptogenesis}.
All our limits apply independently of the Majorana or
Dirac character of the heavy leptons, but in the Majorana case the
restrictions mentioned above must be taken into account.

Finally, let us emphasize that our results are relevant to LHC, since
the production and decay of these new fermions are constrained by the
limits on their mixings that we give here. For production this is
decisive for neutrino singlets, as they can only be produced through
mixing~\cite{Han:2006ip}.
All the other extra leptons can, in addition, be pair produced.
Even if their decays are proportional to the mixings, there is enough
room for the new leptons to decay within the
detector~\cite{delAguila:1989rq}.

The paper is organized as follows. In the next section, after a quick
review of the motivations to consider vector-like leptons, we
enumerate the different possibilities and write down their couplings
to the SM fields. In Section~3 we derive the effective Lagrangian
describing the effect of the new leptons below threshold. We also
describe the 
constraints from FCNC and neutrino masses. In Section~4 we introduce
the observables entering the fit, and present our results for the
different cases. Limits on the mixings are given in the general case
and with the assumption of universality. Section~5 is devoted to a
detailed discussion of the interplay between heavy lepton singlets and
the Higgs mass. Section~6 contains our conclusions, including the
implications of our fits for the observation of heavy 
leptons at large colliders. Finally, the appendix
contains two tables with the experimental and SM values of the 
observables that we use, together with the predictions for two
relevant types of new leptons. 

\section{Extending the Standard Model with vector-like leptons}
Many models of physics beyond the SM include new leptons. Usually,
they are vector-like, i.e. both chiralities
transform in the same way under the SM gauge group. This serves to avoid
constraints from gauge anomalies and also to allow masses above the
electroweak scale without spoiling perturbativity.  By vector-like, we
refer also to Majorana fermions, for which both chiralities are not
independent but related by charge conjugation.
The classical example is SO(10) GUTs, which necessarily contain new
singlets (the right-handed neutrinos). More recent examples 
include models in extra dimensions with leptons propagating in the 
bulk~\cite{xdims} and most Little Higgs models~\cite{littlehiggs}.  
On the other hand, new leptons with masses of the order of 1~TeV and
relatively large mixing with the SM leptons may be observable at
future $e^+\,e^-$ colliders~\cite{del Aguila:2005pf} and even at LHC
in some favourable 
scenarios~\cite{Han:2006ip}. They can also give deviations in
neutrino couplings, which could be measured at
future neutrino experiments~(see for instance
\cite{Adams:2008cm}). Finally, these 
fields can induce lepton FCNC, and in some cases
give mass to the light neutrinos. The current limits on the former,
and the smallness of the latter impose stringent constraints, 
which we discuss in the next section.

It is therefore interesting to study the impact of new vector-like
leptons at the TeV scale on low-energy observables, and the limits that
can be derived on their couplings and masses.
To give sizable contributions to EWPD, the new leptons must mix at
tree level with the SM 
charged leptons and/or neutrinos. This condition and the fact that the
theory must be invariant under the SM gauge group restrict the quantum
numbers of the new particles. All the possibilities are displayed in
Table~\ref{Leptons}, which also settles our notation for the extra
multiplets. 
%
\begin{table}[th]
\begin{center}
{\scriptsize
\begin{tabular}{| c | c c c c c c |} \hline
$\mbox{Leptons}$&$N$&$E$&$\left(\begin{array}{l}N\\E^-\end{array}\right)$&$\left(\begin{array}{l}E^-\\E^{--}\end{array}\right)$&$\left(\begin{array}{l}E^+\\N\\E^-\end{array}\right)$&$\left(\begin{array}{l}N\\E^-\\E^{--}\end{array}\right)$\\
$ $&$ $&$ $&$ $&$ $&$ $&$ $\\[-0.2cm] 
$\mbox{Notation}$&$ $&$ $&$\Delta_{1}$&$\Delta_{3}$&$\Sigma_0$&$\Sigma_1$\\
\hline\hline
$ $&$ $&$ $&$ $&$ $&$ $&$ $\\[-0.1cm]
$SU\left(2\right)_L\otimes U\left(1\right)_Y$&$1_0$&$1_{-1}$&$2_{-\frac 12}$&$2_{-\frac 32}$&$3_0$&$3_{-1}$\\
$ $&$ $&$ $&$ $&$ $&$ $&$ $\\
$ $&$\mbox{Dirac or}$&$\mbox{Dirac}$&$\mbox{Dirac}$&$\mbox{Dirac}$&$\mbox{Dirac or}$&$\mbox{Dirac}$\\[-0.4cm]
$\mbox{Spinor}$&$ $&$ $&$ $&$ $&$ $&$ $\\[-0.05cm]
$ $&$\mbox{Majorana}$&$ $&$ $&$ $&$\mbox{Majorana}$&$ $\\[-0.2cm]
$ $&$ $&$ $&$ $&$ $&$ $&$ $\\
\hline
\end{tabular}}
\caption{Lepton multiplets mixing with the SM leptons through Yukawa
couplings to the SM Higgs. The electric charge is given by $Q=T_3+Y$.} 
\label{Leptons}
\end{center}
\end{table}
%
%
\begin{table}[t]
\begin{center}
{\scriptsize
\begin{tabular}{| c | c c c c c c |} \hline
$ $&$ $&$ $&$ $&$ $&$ $&$ $\\[-0.2cm]
$\overline{\left[L_1\right]}\left[L_2\right] $&$\overline{2_{-\frac 12}}1_0$&$\overline{2_{-\frac 12}}1_{-1}$&$\overline{2_{-\frac 32}}1_{-1}$&$\overline{3_{0}}2_{-\frac 12}$&$\overline{3_{-1}}2_{-\frac 12}$&$\overline{3_{-1}}2_{-\frac 32}$\\[0.2cm]
\hline\hline
$ $&$ $&$ $&$ $&$ $&$ $&$ $\\[-0.1cm]
$\Phi_{L_1 L_2}$&$\tilde \phi$&$\phi$&$\tilde \phi$&$\tilde\phi^\dagger \frac{\sigma_a}{2}$&$\phi^\dagger\frac{\sigma_a}{2}$&$\tilde\phi^\dagger\frac{\sigma_a}{2}$\\[0.15cm]
\hline
\end{tabular}}
\caption{Form of the scalar doublet required to make the operators
$\overline{L_L}\Phi L_R'$, $\overline{L_R}\Phi l_L$ and
$\overline{L_L}\Phi e_R$ gauge invariant, in terms of the quantum
numbers of the leptons appearing in the operator. As usual, $\tilde
\phi= i \sigma_2 \phi^*$ denotes the $Y=-1/2$ doublet.} 
\label{Scalar Form}
\end{center}
\end{table}
%
We consider a 
generic renormalizable extension of the SM including these
fields. After diagonalizing the kinetic and mass matrices of all the
leptons in the theory (before electroweak symmetry breaking), the
Lagrangian of this theory can be split into three pieces:
\begin{equation}
{\cal L}={\cal L}_{\ell}+{\cal L}_h+{\cal L}_{\ell h} .
\end{equation}
${\cal L}_{\ell}$ is the SM Lagrangian and contains only light
fields (with no right-handed neutrinos). We choose a basis in which
the leptonic Yukawa terms are 
diagonal. Then the leptonic sector is given by  
\begin{equation}
{\cal L}_{\ell}\supset\overline{l_L^i}iD\hspace{-0.25cm}/\
l_L^i+\overline{e_R^i}iD\hspace{-0.25cm}/\
e_R^i-\left(\left(\lambda_{e}\right)_i\overline{l_L^i}\ \phi\
e^i_R+\mbox{h.c.}\right) . 
\label{LagSM}
\end{equation}
Here,
$l_L^i=\left(\begin{array}{c}\nu_L^i\\e_L^i\end{array}\right)$
  denotes the left-handed SM doublets, $e^i_R$ denotes the
  right-handed singlets, $\phi$ is the scalar doublet
$\left(\begin{array}{c}\phi^+\\ \phi^0\end{array}\right)$ , and we use 
  lower-case latin letters $i,j$ as family indices. 

${\cal L}_h$ contains the terms involving heavy vector-like leptons
and no SM leptons: 
\begin{equation}
{\cal L}_h=\eta_L\overline{L^I}iD\hspace{-0.25cm}/\ L^I-\eta_LM_I
\overline{L^I}L^I-\left(\left(\lambda_{LL'}\right)_{IJ}\overline{L^I_L}
\Phi_{LL'}L_R^{'J}+\mbox{h.c.}\right)
. 
\label{LagL}
\end{equation}
$L_{L,R}$ stands for the two chiral components of any of the
multiplets in Table \ref{Leptons} while $L$ is the corresponding Dirac
spinor. In the basis we are using, the mass matrices $M$ are diagonal
and real. We also allow for the possibility that $L$ be Majorana when
$L=N$ or $L=\Sigma_0$, and adjust the normalization constants $\eta_L$
with the standard values 1 and $\frac{1}{2}$ for Dirac and Majorana
spinors, respectively. 
The capital latin superindices $I,J$ refer to
the different exotic species with the same quantum numbers. 
Finally, $\Phi_{LL'}$ represents the form of the SM scalar 
doublet needed for gauge invariance of the Yukawa terms, which can be
read from Table~ \ref{Scalar Form}. 

The last piece, ${\cal L}_{\ell h}$, contains all the couplings between
light and heavy fermions, which are of the Yukawa type:
\begin{equation}
{\cal L}_{\ell
h}=-\left(\lambda_{Le}\right)_{Ij}\overline{L^I_L}\Phi_{Le}e_R^j-\left(\lambda_{Ll}\right)_{Ij}\overline{L^I_R}\Phi_{Ll}l_L^j+\mbox{h.c.}
\label{LagSML}
\end{equation}

After electroweak symmetry breaking with 
$\left< \phi\right>=\frac{1}{\sqrt{2}}\left(\begin{array}{c}0\\
v\end{array}\right)$,  
$v=246 \mbox{ GeV}$, mass terms mixing SM and extra
leptons appear. If each SM flavour mixes at most with one extra
lepton, as we shall eventually assume, the diagonalizing $2\times 2$
matrices are given by a mixing $s=\sin \theta$, up to 
phases. We take this mixing to be non-negative, except for
$\Sigma_1$, where we keep a convenient relative minus sign between the
mixing of $\nu_L$ and $e_L$. 
At first order, the mixings are given by ratios of Yukawas $\lambda$
to heavy masses $M$ (times $v$). The precise expressions for the
different  
possible extra leptons are collected in Table~\ref{Mixings}. 
After the diagonalization, the charged and neutral
currents for light and heavy mass eigenstates are written as a
function of the lepton mixings $s$. The strength of the
interactions involving only 
light leptons are modified with respect to the SM ones, correcting
EWPD. This is the subject of this paper. On the other hand, the
very same mixings appear in the charged and neutral currents with one
light and one heavy lepton, which are relevant for the production and
decay of these heavy particles at large colliders. We present our
results in terms of the complete subset of independent charged current
couplings with  
one light and one heavy lepton given in Table~\ref{CCcoup}. As shown
in this table, they turn out to be directly related to the lepton
mixings. For this reason we shall generically use the term ``mixing''
for both $V$ and $s$.
%
\begin{table}[t]
\begin{center}
{\scriptsize
\begin{tabular}{| c | c c c c c c| }\hline
$ $&$ $&$ $&$ $&$ $&$ $&$ $\\[-.2cm]
$\mbox{ }$& $N$&$E$&$\Delta_1$&$\Delta_3$&$\Sigma_0$&$\Sigma_1$\\[.1cm]
\hline\hline 
$ $&$ $&$ $&$ $&$ $&$ $&$ $\\[-.15cm]
$s_L^\nu $&
$\left|\frac{\lambda_{Nl}v}{\sqrt{2}M_N}\right| $&$- $&$- $&$- $&$\left|\frac{\lambda_{\Sigma_0 l}v}{2\sqrt{2}M_{\Sigma_0}}\right| $&$-\sqrt{2}s_L^e$\\[0.2cm]
$s_L^e$&
$- $&$\left|\frac{\lambda_{El}v}{\sqrt{2}M_E}\right| $&$\mbox{negligible}$&$\mbox{negligible}$&$\sqrt{2}s_L^\nu $&$\left|\frac{\lambda_{\Sigma_1 l}v}{2\sqrt{2}M_{\Sigma_1}}\right|  $\\[0.2cm]
$s_R^e $&
$- $&$\mbox{negligible}$&$\left|\frac{\lambda_{\Delta_1 e}v}{\sqrt{2}M_{\Delta_1}}\right| $&$\left|\frac{\lambda_{\Delta_3 e}v}{\sqrt{2}M_{\Delta_3}}\right| $&$\mbox{negligible}$&$\mbox{negligible}$\\[0.2cm]\hline
\end{tabular}}
\caption{First order expressions in $\frac{\lambda v}{M}$ of the
mixing between one SM lepton of a given
flavour and one extra lepton. Family indices are 
implicit and ``negligible'' stands for higher order contributions.}
\label{Mixings}
\end{center}
\end{table}
%
%
\begin{table}[t]
\begin{center}
{\scriptsize
\begin{tabular}{| c | c c c c c c| }\hline
$ $&$ $&$ $&$ $&$ $&$ $&$ $\\[-.2cm]
$\mbox{ }$& $N$&$E$&$\Delta_1$&$\Delta_3$&$\Sigma_0$&$\Sigma_1$\\[.1cm]
\hline\hline 
$ $&$ $&$ $&$ $&$ $&$ $&$ $\\[-.15cm]
$\overline{f_A}\gamma^\mu f_{A}'$&
$\overline{e_L^-}\gamma^\mu N_L$&$\overline{E_L^{-}}\gamma^\mu \nu_L$&$\overline{e_R^{-}}\gamma^\mu N_R$&$\overline{E_R^{--}}\gamma^\mu e_R^{-}$&$\overline{e_L^{-}}\gamma^\mu N_L$&$\overline{E_L^{-}}\gamma^\mu \nu_L$\\[.1cm]

$\left|V^{ff'}_{A}\right| $&
$s_L^\nu$&$s_L^e $&$s_R^e $&$ s_R^e$&$s_L^\nu $&$s_L^e $\\[0.2cm]\hline
\end{tabular}}
\caption{Resulting first order expressions of a complete subset of
  independent charged current couplings
  $-\frac{g}{\sqrt{2}}V_{A}^{ff'}W_\mu^- \overline{f_{A}}\gamma^\mu f_{A}'$, $A=L,R$, as a functions of the lepton mixings.}
\label{CCcoup}
\end{center}
\end{table}
%

\section{Effective Lagrangian}
As we are interested in the effects of the heavy particles at energies
much smaller than their masses, we can integrate them out and use the
resulting effective Lagrangian. This is completely equivalent, for our
purposes, to diagonalizing the mass matrices to first order and using
the resulting charged and neutral couplings for light
fields. Nevertheless, we find it interesting to write down the
completely-gauge-invariant induced operators and their coefficients
before electroweak symmetry breakdown.
In particular, this may be useful 
to compare with other new physics effects in EWPD.
Because the heavy leptons are
vector-like, they decouple in the limit when their mass goes to
infinity. Therefore,  
we expand the effective Lagrangian as
\begin{equation}
{\cal L}_\mathrm{eff}={\cal L}_4+\frac{1}{\Lambda}{\cal
  L}_5+\frac{1}{\Lambda^2}{\cal L}_6+\dots 
\label{Leff}
\end{equation}
where each ${\cal L}_d$ contains gauge-invariant local operators of
canonical dimension $d$, and the scale $\Lambda$ is equal to the
mass $M$ of the lightest new lepton. The operators in ${\cal L}_d$ give
contributions of order $(E/\Lambda)^{d-4}$ to observables, with $E$
the typical energy of the processes involved or the vacuum expectation
value $v$ of the scalar field. We expect the terms of dimension $d>6$
to give small corrections compared to the experimental precision of 
current data, so we neglect them in the fits. Our results will be
consistent with this approximation. 

In $\mathcal{L}_5$ there is only one operator:
\begin{equation}
{\cal L}_5=(\alpha_5)_{ij} \overline{(l_L^i)^c}\tilde \phi ^* \tilde
\phi^\dagger l_L^j+\mbox{h.c.} 
\label{L5} 
\end{equation}
This is the lepton number violating Weinberg
operator~\cite{Weinberg:1979sa}, which after 
electroweak symmetry breaking gives masses to the light neutrinos,
$m_\nu=-v^2\alpha_5/\Lambda$, with 
$(\alpha_5)_{ee}$ contributing also to neutrinoless double $\beta$ decay.   
This operator can originate from 
Majorana terms in \refeq{LagL}, which are possible only for extra
singlets or triplets of zero hypercharge. The value of the coefficient
$\alpha_5$ is given in Table~ \ref{Coeff1}. 
The fact that neutrino masses are tiny, and the strict bounds on
neutrino double $\beta$ decay, are usually explained by a large
scale $\Lambda$. However, we want to keep the
scale $\Lambda$ near the TeV range to have non-negligible effects
from ${\cal L}_6$. Then we need to assume that some mechanism in the
high energy model keeps the coefficient $(\alpha_5)_{ij}$
very small. A natural way to achieve this in any model is to 
implement lepton number conservation, up to possible
breaking terms with adimensional coefficients $\alpha_5$ smaller than
$10^{-11}$~\cite{delAguila:2007ap}. This scenario is stable under
quantum corrections and is realized in models in which the heavy
fermions are of Dirac type~\cite{Kersten:2007vk}. Unnatural
cancellations are also possible~\cite{Ingelman:1993ve}. 

At order $1/\Lambda^2$, we find 
\begin{equation}
\begin{split}
{\cal L}_6&=\left(\alpha_{\phi l}^{(1)}\right)_{ij} \left(\phi^\dagger
iD_\mu \phi\right)\left(\overline{l_L^i}\gamma^\mu l_L^j\right) +
\left(\alpha_{\phi l}^{(3)}\right)_{ij} \left(\phi^\dagger
i\sigma_aD_\mu \phi\right)\left(\overline{l_L^i}\sigma_a\gamma^\mu
l_L^j\right)+\\ 
&+\left(\alpha_{\phi e}^{(1)}\right)_{ij}\left(\phi^\dagger iD_\mu
\phi\right)\left(\overline{e_R^i}\gamma^\mu
e_R^j\right)+\left(\alpha_{e\phi}^{ }\right)_{ij}
\left(\phi^\dagger\phi\right)\overline{l_L^i}\phi e_R^j+\mbox{h.c.} . 
\label{L6} 
\end{split} 
\end{equation}
We have made field redefinitions to write the operators in the basis
of Buchmuller and Wyler~\cite{Buchmuller:1985jz}, and follow the notation in this
reference. 
The values of the coefficients of the operators are given in Tables
\ref{Coeff1} and \ref{Coeff2}. These results parallel the ones
obtained and discussed in~\cite{delAguila:2000rc} for extra quarks. After
electroweak symmetry breaking these operators modify the neutral
current and charged current couplings of leptons: 
\begin{equation}
\begin{split}
\delta g_L^\nu&=\frac 14\left(-\alpha_{\phi l}^{(1)}+\alpha_{\phi l}^{(3)}+\mbox{h.c.}\right)\frac{v^2}{\Lambda^2}\\
\delta g_L^e&=-\frac 14\left(\alpha_{\phi l}^{(1)}+\alpha_{\phi l}^{(3)}+\mbox{h.c.}\right)\frac{v^2}{\Lambda^2}\\
\delta g_R^e&=-\frac 14\left(\alpha_{\phi e}^{(1)}+\mbox{h.c.}\right)\frac{v^2}{\Lambda^2}\\
\delta V_L^{e\nu}&=\left(\alpha_{\phi l}^{(3)}\right)^\dagger
\frac{v^2}{\Lambda^2} \, .  
\end{split}
\label{Dcorrect}
\end{equation}
Here, the $\delta g$ and $\delta V$ are in principle general matrices.
The charged lepton masses 
and their Yukawa couplings to the Higgs are also modified, but these
changes can be absorbed into the observed charged lepton
masses. Moreover, neglecting tiny irrelevant contributions from
neutrino masses, we can re-diagonalize the mass matrix with bi-unitary
transformations that 
do not introduce further changes in the neutral and charged currents
to order $1/\Lambda^2$.
So, the operator  
$\left(\phi^\dagger\phi\right)\overline{l_L^i}\phi e_R^j$ 
and the corresponding coefficients $\alpha_{e\phi}$ in
Tables~\ref{Coeff1} and \ref{Coeff2} do   
not contribute to our fits. 
Observe also in Table~\ref{Coeff1} that the combinations of
Yukawa couplings entering $\alpha_{\phi l,e}^{(1,3)}$ is different
from the ones in $\alpha_5$, so that it is perfectly possible to have
finite $\alpha_{\phi l,e}^{(1,3)}$ and vanishing
$\alpha_5$ simultaneously, even for $N$ and $\Sigma_0$
multiplets~\cite{delAguila:2007ap}. 
%
\begin{table}[th]
\begin{center}
{\scriptsize
\begin{tabular}{| l | c c c c c |} \hline
$\ L\ $&$\frac{\alpha_5}{\Lambda}$&$\frac{\alpha_{\phi l}^{(1)}}{\Lambda^2}$&$\frac{\alpha_{\phi l}^{(3)}}{\Lambda^2}$&$\frac{\alpha_{\phi e}^{(1)}}{\Lambda^2}$&$\frac{\alpha_{e\phi}}{\Lambda^2}$\\
$ $&$ $&$ $&$ $&$ $&$ $\\[-0.2cm]
\hline\hline
$ $&$ $&$ $&$ $&$ $&$ $\\[-0.3cm]
$\ N\ $&$\frac 12 \lambda^T_{Nl}M_N^{-1}\lambda_{Nl}$&$\frac 14 \lambda^\dagger_{Nl}M_N^{-2}\lambda_{Nl}$&$-\frac{\alpha_{\phi l}^{(1)}}{\Lambda^2}$&$-$&$-$\\
$ $&$ $&$ $&$ $&$ $&$ $\\[-0.3cm]
$ $&$ $&$ $&$ $&$ $&$ $\\[-0.3cm]
$\ E\ $&$-$&$-\frac 14 \lambda^\dagger_{El}M_E^{-2}\lambda_{El}$&$\frac{\alpha_{\phi l}^{(1)}}{\Lambda^2}$&$-$&$-2\frac{\alpha_{\phi l}^{(1)}}{\Lambda^2}\lambda_e$\\
$ $&$ $&$ $&$ $&$ $&$ $\\[-0.3cm]
$ $&$ $&$ $&$ $&$ $&$ $\\[-0.3cm]
$\ \Delta_1\ $&$-$&$-$&$-$&$\frac 12 \lambda^\dagger_{\Delta_{1}e}M_{\Delta_1}^{-2}\lambda_{\Delta_{1}e}$&$\lambda_e\frac{\alpha_{\phi e}^{(1)}}{\Lambda^2}$\\
$ $&$ $&$ $&$ $&$ $&$ $\\[-0.3cm]
$ $&$ $&$ $&$ $&$ $&$ $\\[-0.3cm]
$\ \Delta_3\ $&$-$&$-$&$-$&$-\frac 12 \lambda^\dagger_{\Delta_{3}e}M_{\Delta_3}^{-2}\lambda_{\Delta_{3}e}$&$-\lambda_e\frac{\alpha_{\phi e}^{(1)}}{\Lambda^2}$\\
$ $&$ $&$ $&$ $&$ $&$ $\\[-0.3cm]
$ $&$ $&$ $&$ $&$ $&$ $\\[-0.3cm]
$\ \Sigma_0\ $&$\frac 18 \lambda^T_{\Sigma_0l}M_{\Sigma_0}^{-1}\lambda_{\Sigma_0l}$&$\frac {3}{16} \lambda^\dagger_{\Sigma_0l}M_{\Sigma_0}^{-2}\lambda_{\Sigma_0l}$&$\frac {1}{3}\frac{\alpha_{\phi l}^{(1)}}{\Lambda^2}$&$-$&$\frac 43\frac{\alpha_{\phi l}^{(1)}}{\Lambda^2}\lambda_e$\\
$ $&$ $&$ $&$ $&$ $&$ $\\[-0.3cm]
$ $&$ $&$ $&$ $&$ $&$ $\\[-0.3cm]
$\ \Sigma_1\ $&$-$&$-\frac {3}{16}\lambda^\dagger_{\Sigma_1l}M_{\Sigma_1}^{-2}\lambda_{\Sigma_1l}$&$-\frac {1}{3}\frac{\alpha_{\phi l}^{(1)}}{\Lambda^2}$&$-$&$-\frac 23 \frac{\alpha_{\phi l}^{(1)}}{\Lambda^2}\lambda_e$\\
$ $&$ $&$ $&$ $&$ $&$ $\\
\hline
\end{tabular}} 
\caption{Coefficients of the operators arising from the integration of
heavy leptons. The dimension five operator entry,
$\frac{\alpha_5}{\Lambda}$, only appears when the singlet $N$ and/or
the triplet $\Sigma_0$ are Majorana fermions.} 
\label{Coeff1}
\end{center}
\end{table}
\begin{table}[th]
\begin{center}
{\scriptsize
\begin{tabular}{| c | c |} \hline
$ $&$ $\\[-0.25cm]
$L_1,L_2 $&$\frac{\alpha_{e\phi}}{\Lambda^2}$\\
$ $&$ $\\[-0.3cm]
\hline\hline
$ $&$ $\\[-0.2cm]
$E,\Delta_1$&$\lambda_{El}^\dagger M_E^{-1}
\lambda_{E\Delta_{1}}M_{\Delta_1}^{-1}\lambda_{\Delta_{1}e}$\\ 
$ $&$ $\\[-0.1cm]
$E,\Delta_3$&$\lambda_{El}^\dagger M_E^{-1}
\lambda_{E\Delta_{3}}M_{\Delta_3}^{-1}\lambda_{\Delta_{3}e}$\\ 
$ $&$ $\\[-0.1cm]
$\Delta_1,\Sigma_0$&$\frac 12 \lambda_{\Sigma_{0}l}^\dagger
M_{\Sigma_0}^{-1}
\lambda_{\Sigma_0\Delta_{1}}M_{\Delta_1}^{-1}\lambda_{\Delta_1 e}$\\  
$ $&$ $\\[-0.1cm]
$\Delta_1,\Sigma_1$&$\frac 14 \lambda_{\Sigma_{1}l}^\dagger
M_{\Sigma_1}^{-1}
\lambda_{\Sigma_{1}\Delta_1}M_{\Delta_1}^{-1}\lambda_{\Delta_1 e}$\\  
$ $&$ $\\[-0.1cm]
$\Delta_3,\Sigma_1$&$-\frac 14 \lambda_{\Sigma_{1}l}^\dagger
M_{\Sigma_1}^{-1} \lambda_{\Sigma_{1}\Delta_3}M_{\Delta_3}^{-1}
\lambda_{\Delta_3 e}$\\[0.2cm]\hline   
\end{tabular}}
\caption{Combined contribution to $\alpha_{e\phi}$ from the
  simultaneous integration of different mixed multiplets. Even if the
  corresponding operator does not affect our fits, we include the
  values of the coefficient for completeness.}
\label{Coeff2}
\end{center}
\end{table}
%

On the other hand, the off-diagonal elements of the coefficient
matrices $\alpha_{\phi l,e}^{(1,3)}$ induce leptonic FCNC. The current
experimental limits on rare processes like $\mu
\rightarrow e \gamma$ and $\mu \rightarrow e e e\,$ imply that these
off-diagonal coefficients are small~\cite{Abada:2007ux}.  As can be seen 
from Table~\ref{Coeff1}, this requires that each new fermion multiplet
mixes mostly with only one of the known lepton flavours. This
pattern of mixings is automatic with the extra assumption of an
(approximate) conservation of individual lepton number.

\section{Global fit}
\label{section_globalfit}
We have performed global fits to the existing EWPD to confront the
hypothesis of new leptons with the SM, and to constrain the new parameters
(lepton mixings). 
In the appendix, we show in Tables~\ref{Exp-SM} and~\ref{Exp-SM-LU}
the observables that enter our 
fits, together with their current experimental values and the SM
predictions.
We do not include data at higher
energies from LEP~2 because they do not change significantly the
fits. The reason is that the Z-pole observables have better precision
and constrain strongly all the new parameters in the model, i.e. no
independent parameters enter the LEP~2 data. This can be understood
by the fact 
that the new leptons change only the trilinear couplings, and do not
generate four-fermion operators in the effective Lagrangian.

With the experimental data we construct the $\chi^2$ function to be minimized:
\begin{equation}
\chi^2\left(\theta\right)=
\left[\mbox{O}_\mathrm{exp}-\mbox{O}_\mathrm{th}\left(\theta\right)\right]^T \,
U_\mathrm{exp}^{-1} \,
\left[\mbox{O}_\mathrm{exp}-\mbox{O}_\mathrm{th}\left(\theta\right)\right] 
, 
\end{equation}
where $(U_\mathrm{exp})_{ij}=\sigma_i\rho_{ij}\sigma_j$ is the
covariance matrix,  
with $\sigma$ the experimental errors and 
$\rho$ the correlation matrix, and $\theta$ are the free
parameters. In $U_\mathrm{exp}$ we include both statistical and systematic
errors. 
$\mbox{O}_\mathrm{exp}$ are the experimental values of the (pseudo)
observables and  
$\mbox{O}_\mathrm{th}\left(\theta\right)$ contains the theoretical
predictions obtained from $\mathcal{L}_\mathrm{eff}$ and expressed in
terms of the parameters of the original model (SM + new leptons). The
good agreement of the SM with the experimental data 
allows us to consider only the corrections coming from the
interference between the SM and the new pieces in the effective
Lagrangian. This means that we calculate only
tree-level contributions from new physics, and linearize the values of
the observables in $v^2/\Lambda^2$. We use ZFITTER~6.42~\cite{Arbuzov:2005ma}
to compute the SM predictions at the quantum level. 

Within our approximations, the new free parameters of the model always
enter the fit as ratios of Yukawa couplings to heavy masses, corresponding 
to the mixing between light and heavy particles as explained above and 
gathered in Table \ref{Mixings}. 
We present our results in function of the equivalent charged current 
couplings $V$ in Table \ref{CCcoup}.
The fits constrain only the magnitudes $\left|V\right|$.
The new leptons can modify the observables in two ways. First, they
can give direct contributions to the processes relevant to a given
observable. Second, they can contribute to the processes from which
the input parameters are extracted. This changes the relation between
the measured values and the SM parameters, and results in indirect
corrections to all the observables. 

The free parameters in the fits are $\Delta
\alpha_\mathrm{had}^{(5)}\left(M_Z^2\right)$,
$\alpha_S\left(M_Z^2\right)$, $M_Z$, $m_t$, $M_H$ 
and the mixings of the new leptons. Note that the first four
parameters are to a great extent determined by the corresponding
experimental measurements\footnote{We can neglect the effect of the
heavy leptons on these measurements. In
particular, for $\alpha_S$ we 
take the world average in~\cite{Yao:2006px}. Even if this average
includes the 
SM fit to EWPD as an input, the central value will 
  not be changed significantly by the presence of new leptons, and we
  use the most conservative error interval given in that
  reference.}. 
Therefore, only $M_H$ and the mixings can vary
significantly and we will give the results in terms of these two
parameters. Furthermore, we make use of the information from
direct Higgs searches at LEP by imposing a sharp lower cut-off on the
Higgs mass, $M_H\geq 114.4$~GeV. This is a good approximation to the
more precise treatment proposed in~\cite{delAguila:1992ba}.

The minimization of $\chi^2$ and the calculation of the confidence
regions are performed by scanning over the parameter
space\footnote{In practice, for the reasons discussed above, we
  restrict the parameters $\Delta 
\alpha_\mathrm{had}^{(5)}\left(M_Z^2\right)$,
$\alpha_S\left(M_Z^2\right)$, $M_Z$, $m_t$ to $1\, \sigma$ intervals
around their SM value.} and accepting or rejecting points according to
their probability. 
The plots are obtained from the actual sets
of points, keeping only the points within the 90\% probability regions
and performing a coarse graining to lower the size of the figures.

\subsection{Numerical results}
%
\begin{table}[th]
\begin{center}
{\scriptsize
\begin{tabular}{ |c c | c c c c c c |}\hline
$ $&$ $&$ $&$ $&$ $&$ $&$ $&$ $\\[-0.1cm]
$ $&$ $&$ $&$ $&$-\Delta
  \chi^2_{min}$&$(\chi^2_{min}/\mbox{d.o.f.})$&$ $&$ $\\[.1cm] 
$ $&$ $&$ $&$ $&$ $&$ $&$ $&$ $\\[-0.3cm]
$\mbox{Coupling}$&
  $n^{new}_{par}$&$N$&$E$&$\Delta_1$&$\Delta_3$&$\Sigma_0$&$\Sigma_1$\\[.2cm] 
\hline\hline
$ $&$ $&$ $&$ $&$ $&$ $&$ $&$ $\\[-0.2cm]
$\mbox{General}$&
$3$&$1.5\,(1.57 )$&$ 0.5\,(1.61)$&$1.9 \,(1.56 )$&$1.5 \,(1.57 )$&$1.3
  \,(1.58 )$&$0\,(1.63 )$\\[.1cm]  
$\mbox{Universal}$&
$1$&$1.0\,(1.44 )$&$0\,(1.49 )$&$0\,(1.49 )$&$0.3\,(1.48
  )$&$0.7\,(1.46 )$&$0\,(1.49)$\\[.1cm]  
$\mbox{Only with }e$&
$1$&$0.8 \,(1.49 )$&$0 \,(1.51 )$&$0\, (1.51 )$&$0.7 \,(1.49 )$&$1.0
  \,(1.48 )$&$0\,(1.51 )$\\[.1cm]  
$\mbox{Only with }\mu$&
$1$&$1.0 \,(1.48 )$&$0.5 \,(1.50 )$&$1.9 \,(1.45 )$&$0 \,(1.51 )$&$0
\,(1.51 )$&$0 \,(1.51 )$\\[.1cm] 
$\mbox{Only with }\tau$&
$1$&$1.0 \,(1.48 )$&$0 \,(1.51 )$&$0 \,(1.51 )$&$0.6 \,(1.49 )$&$0.2
\,(1.51 )$&$0 \,(1.51 )$\\[.1cm] 
\hline
\end{tabular}}
\caption{Decrease in $\chi^2_{min}$ with respect to the SM minimum,
  $\chi^2_{SM}=43.92$ ($\chi^2_{SM}=29.82$ with lepton universality),
  obtained by 
  adding to the SM the different leptons. The number of degrees of
  freedom is obtained as ${\cal{N}}-5-n^{new}_{par}$, where
  $n^{new}_{par}$ is the number of independent lepton mixings and
  ${\cal{N}}=35$ is the number of observables (${\cal{N}}=26$ for the
  universal case). In parenthesis we write the value of
  $\chi^2_{min}/\mbox{d.o.f.}$, which for the SM is 1.46 (1.42 with
  lepton universality).}
\label{LeptFits}
\end{center}
\end{table}
%

In Table~\ref{LeptFits} we show the improvements $-\Delta
\chi_{min}^2$ with respect to the SM minimum (consistent with $M_H\geq
114.4~\mathrm{GeV}$), when we add independently
one kind of new lepton at a time. We have also performed a general fit
including all possible heavy leptons, but there is no further
significant improvement and we do not show the result here. 
We distinguish different scenarios depending on how we choose the couplings
of the new leptons to the SM fields. We have considered the following
cases:
\begin{itemize}
\item A single new lepton coupled only to one of the three SM families
  (``Only with $e$, $\mu$ or $\tau$'').
\item Three leptons, each coupled to one (different) SM family with
  independent couplings (``General'').
\item Three leptons, each coupled to one (different) SM family with the
  same coupling (``Universal''). 
\end{itemize}
The universal case requires an extra assumption. When we do
the fit with universal couplings, we use this assumption of
universality also in the experimental measurements.
This implies that
the set of data is different and hence the comparison with the other
fits is not direct. The (pseudo) observables included in the fit for
this case, with their current experimental values and
the SM predictions, are collected in Table~\ref{Exp-SM-LU} in
the appendix. 

We see that there are mild improvements with respect to
the SM $\chi^2$ for singlets $N_{\mu,\tau}$, doublets
$(\Delta_1)_\mu$ and triplets $(\Sigma_0)_e$, and also for universal
singlets $N$. In all 
the other cases the $\chi^2$ is lowered by less than one unit. The only
fit with $\chi^2/\mathrm{d.o.f.}$ smaller than in the SM
is obtained for the SM-like doublet coupled to the second family,
$(\Delta_1)_\mu$. Even if the improvements are marginal at best, it is
interesting that in some cases the minima occur for significant values
of the mixings, as can be seen in Table~\ref{LeptLimits}. Let us also
mention the biggest changes in individual 
observables at the global minima. First, $\sigma_H^0$ (with a 1.7 pull
in the SM) is improved in several cases, up to a pull of $0.8$ for the
singlet $N_\tau$. The pull in the SLD 
asymmetry $A_e$ is lowered from 2.0 to 1.7 for singlets $N_{e,\mu}$,
but at the price of increasing the $A_{FB}^{0,b}$ anomaly from 2.6 to
2.8. The NuTeV anomaly is reduced only for universal triplets
$\Sigma_0$, and only from 2.8 to 2.6. Finally, $(\Delta_1)_\mu$,
reduces the pull in $R_\mu^0$ from 1.4 to 0.1. In Tables~\ref{Exp-SM}
and~\ref{Exp-SM-LU} we give, together with the experimental and SM
values, the best-fit values for our set of observables in the
extensions with a doublet $(\Delta_1)_\mu$ and with a universal singlet $N$,
respectively.
%
\begin{table}[ht]
\begin{center}
{\scriptsize
\begin{tabular}{| c c | c c c c c c |}\hline
$ $&$ $&$ $&$ $&$ $&$ $&$ $&$ $\\[-.2cm]
$\mbox{Coupling}$&$\mbox{ }$& $N$&$E$&$\Delta_1$&$\Delta_3$&$\Sigma_0$&$\Sigma_1$\\[.2cm]
\hline\hline
$ $&$ $&$ $&$ $&$ $&$ $&$ $&$ $\\[-.25cm]
$\mbox{Only with }e$&$\left|V\right|< $&
$0.055 $&$0.018 $&$0.018 $&$0.025 $&$0.019 $&$0.013 $\\[.1cm]
$ $&$\left|V_{\mathrm{min}}\right| =$&
$0.035 $&$0 $&$0 $&$0.018 $&$0.014 $&$0 $\\[.3cm]
$ $&$ $&$ $&$ $&$ $&$ $&$ $&$ $\\[-.2cm]
$\mbox{Only with }\mu$&$\left|V\right|< $&
$0.057 $&$0.034 $&$0.045 $&$0.024 $&$0.017 $&$0.022 $\\[.1cm] 
$ $&$\left|V_{\mathrm{min}}\right| =$&
$0.036 $&$0.020 $&$0.035 $&$0 $&$0 $&$0 $\\[.3cm] 
$ $&$ $&$ $&$ $&$ $&$ $&$ $&$ $\\[-.2cm]
$\mbox{Only with }\tau$&$\left|V\right|<$&
$0.079 $&$0.030 $&$0.030 $&$0.042 $&$0.027 $&$0.026 $\\[.1cm]
$ $&$\left|V_{\mathrm{min}}\right| =$&
$0.057 $&$0 $&$0 $&$0.028 $&$0.015 $&$0 $\\[.3cm]
$ $&$ $&$ $&$ $&$ $&$ $&$ $&$ $\\[-.2cm] 
$\mbox{Universal} $&$\left|V\right|<$&
$0.038 $&$0.018 $&$0.019 $&$0.022 $&$0.016 $&$0.011 $\\[.1cm]
$ $&$\left|V_{\mathrm{min}}\right| =$&
$0.025 $&$0 $&$0 $&$0.014 $&$0.012 $&$0 $\\[.1cm]
\hline 
\end{tabular}}
\caption{Upper limit at 90 $\%$ C.L. on the absolute value of the
  mixings in Table~\ref{CCcoup} and their value at 
  the minimum. The first
three rows are obtained by coupling each new lepton with only one SM
family. The last one corresponds to the case of lepton
universality. All numbers are computed assuming $M_H\geq
114.4~\mbox{GeV}$.}  
\label{LeptLimits}
\end{center}
\end{table}
%
%
\begin{table}[ht]
\begin{center}
{\scriptsize
\begin{tabular}{| c c | c c c c c c |}\hline
$ $&$ $&$ $&$ $&$ $&$ $&$ $&$ $\\[-.2cm]
$\mbox{Couplings}$&$\mbox{ }$&$N$&$E$&$\Delta_1$&$\Delta_3$&$\Sigma_0$&$\Sigma_1$\\[.2cm]
\hline\hline
$ $&$ $&$ $&$ $&$ $&$ $&$ $&$ $\\[-.25cm]
$\mbox{Only with }e$&$M_H\left[\mbox{GeV}\right]<$&
$259 $&$166 $&$168 $&$162 $&$168 $&$163 $\\[.1cm]
$ $&$ $&$ $&$ $&$ $&$ $&$ $&$ $\\[-.1cm]
$\mbox{Only with }\mu $&$M_H\left[\mbox{GeV}\right]<$&
$267 $&$187 $&$167 $&$165 $&$163 $&$162 $\\[.1cm]
$ $&$ $&$ $&$ $&$ $&$ $&$ $&$ $\\[-.1cm]
$\mbox{Only with }\tau $&$M_H\left[\mbox{GeV}\right]<$&
$164 $&$165 $&$167 $&$164 $&$167 $&$166 $\\[.1cm]
$ $&$ $&$ $&$ $&$ $&$ $&$ $&$ $\\[-.1cm]
$\mbox{Universal} $&$M_H\left[\mbox{GeV}\right]<$&
$253 $&$171 $&$170 $&$163 $&$166 $&$164 $\\[.1cm]
\hline
\end{tabular}}
\caption{Upper limit at 90 $\%$ C.L. on the Higgs mass (in $\mathrm{ GeV}$).
 The first 
three rows are obtained by coupling each new lepton with only one SM
family. The last one correspond to the case of lepton
universality. All numbers are computed assuming $M_H\geq
114.4~\mbox{GeV}$.} 
\label{MHLimits}
\end{center}
\end{table}
%

From the fits, we can also extract limits on the values of
the mixings $V$ and $s$ in Tables~\ref{Mixings} and~\ref{CCcoup},
respectively.   
We give the 90\%
C.L. upper bounds on the absolute value of $V$ in
Table~\ref{LeptLimits}. We stress again that 
these limits incorporate the information from the 
direct Higgs searches. 
%
\begin{figure}[t]
\includegraphics[width=7cm]{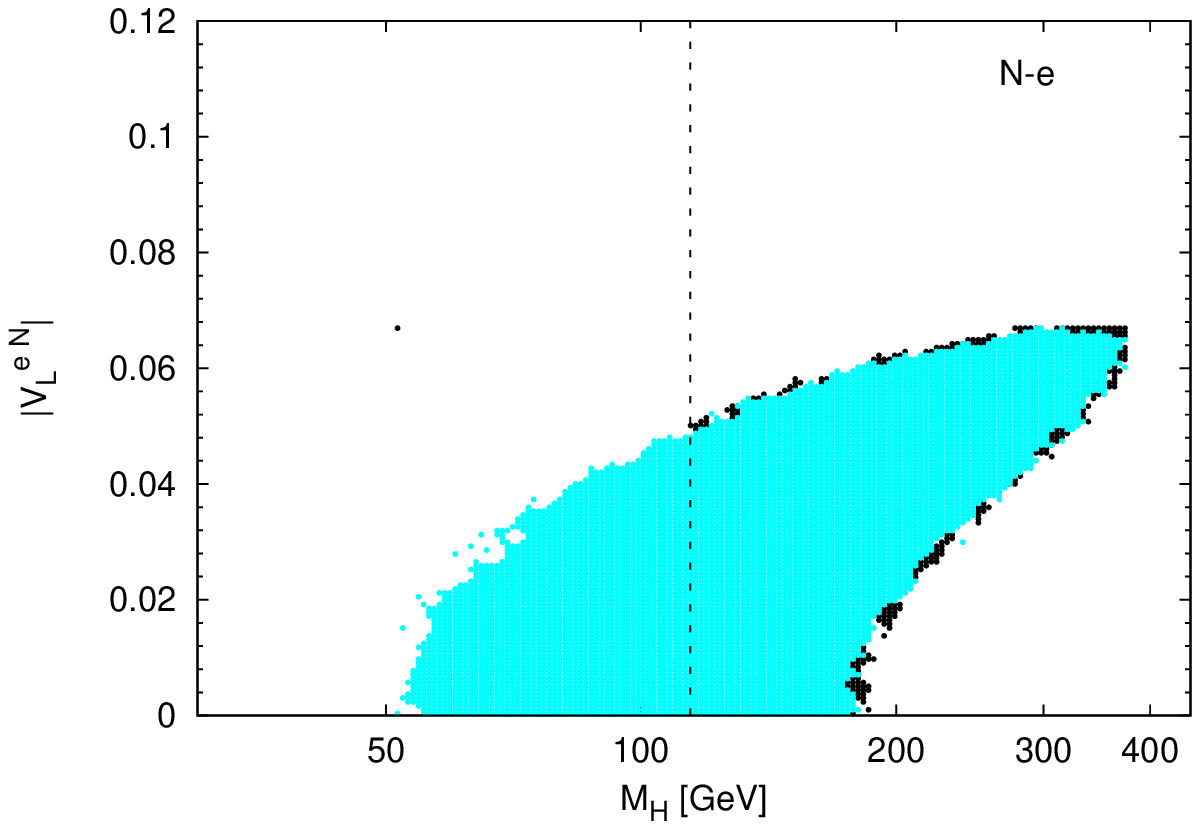}
\includegraphics[width=7cm]{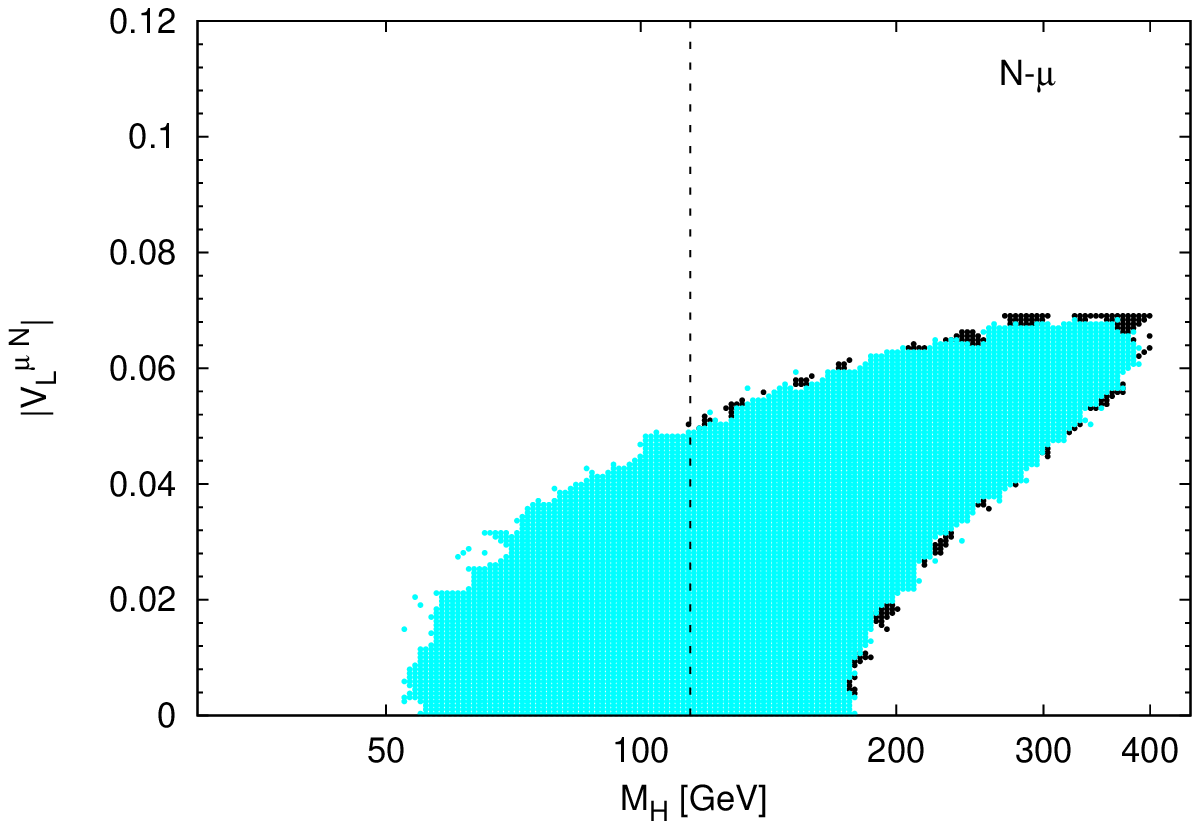}
\includegraphics[width=7cm]{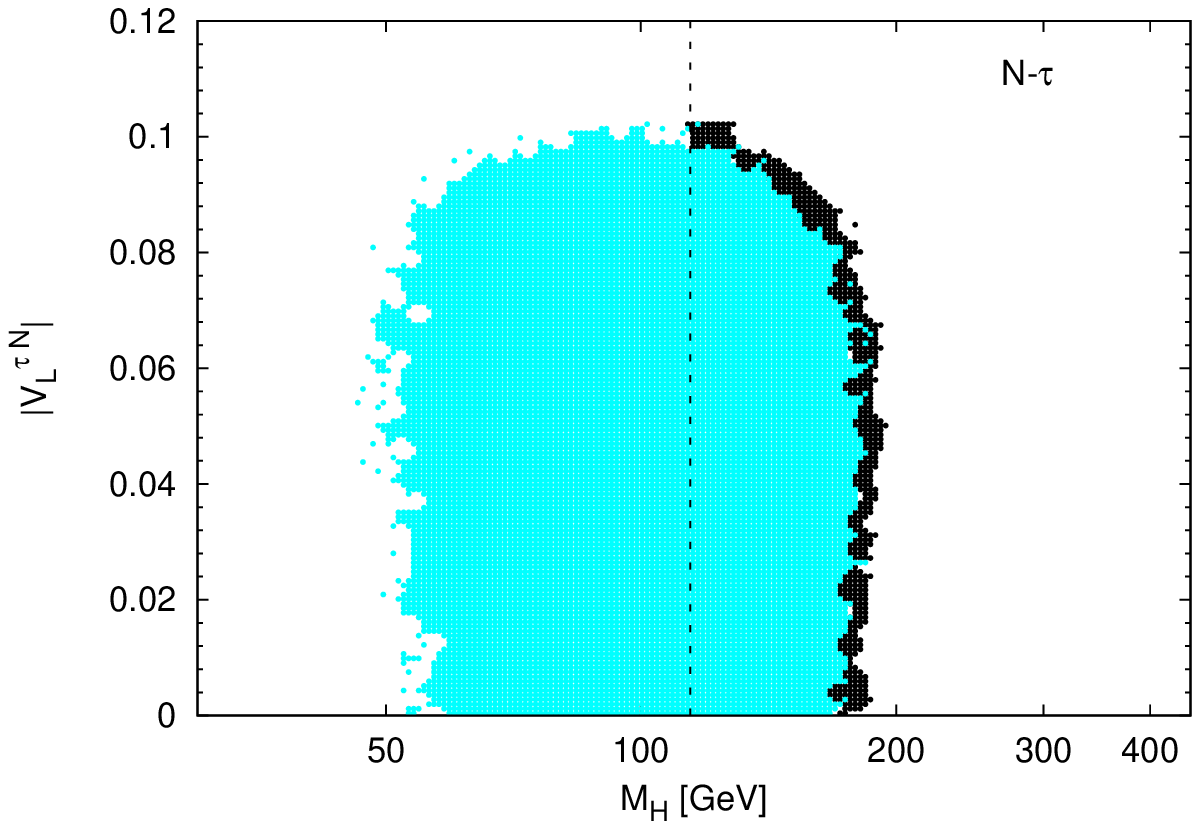}
\includegraphics[width=7cm]{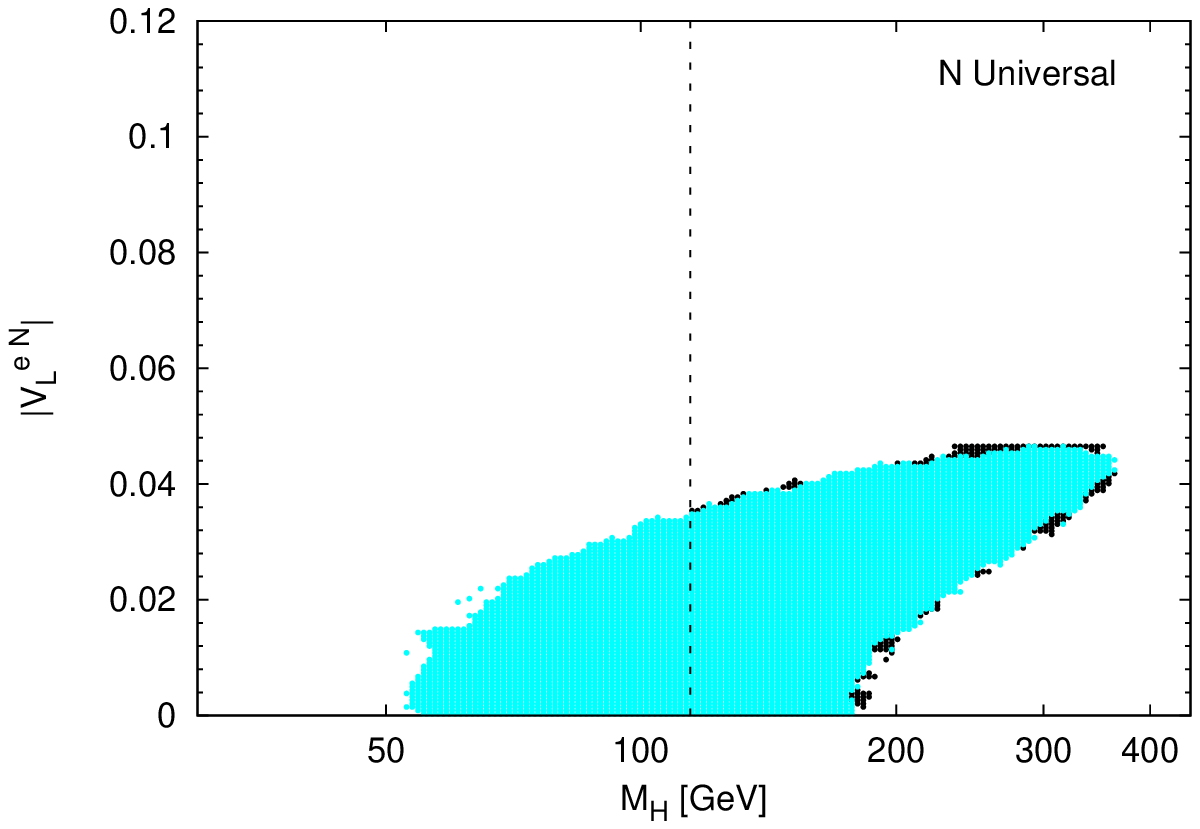}
\caption{90$\%$ confidence region in the $\left|V_L^{eN}\right|-M_H$
parameter space for the $N$ singlet coupled to the first,
second and third family, respectively.  The last plot corresponds
to the universal case. In all cases the extension of the 90$\%$
confidence region with the cut $M_H\geq 114.4 \mbox{ GeV}$ (represented by
the vertical dashed line) is shown in black. 
\label{LimitplotsN}}  
\end{figure}
%
%
\begin{figure}[t]
\includegraphics[width=7cm]{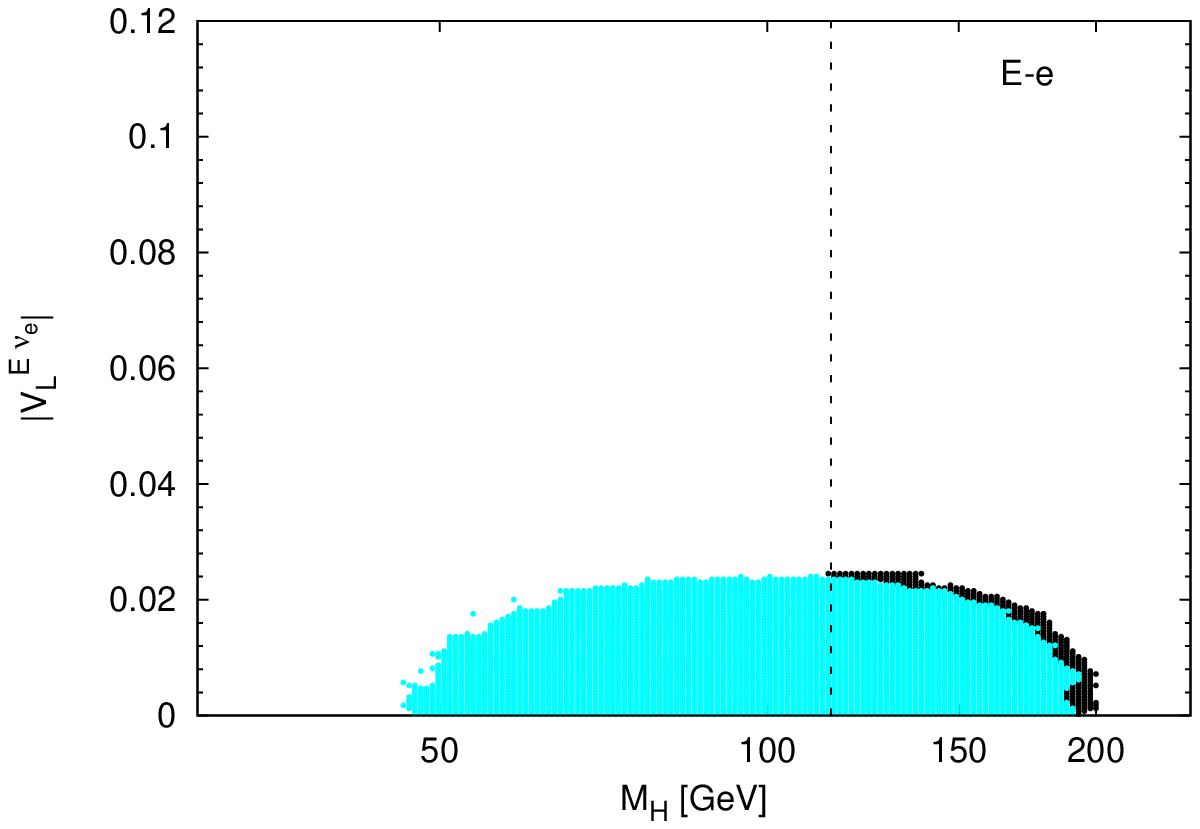}
\includegraphics[width=7cm]{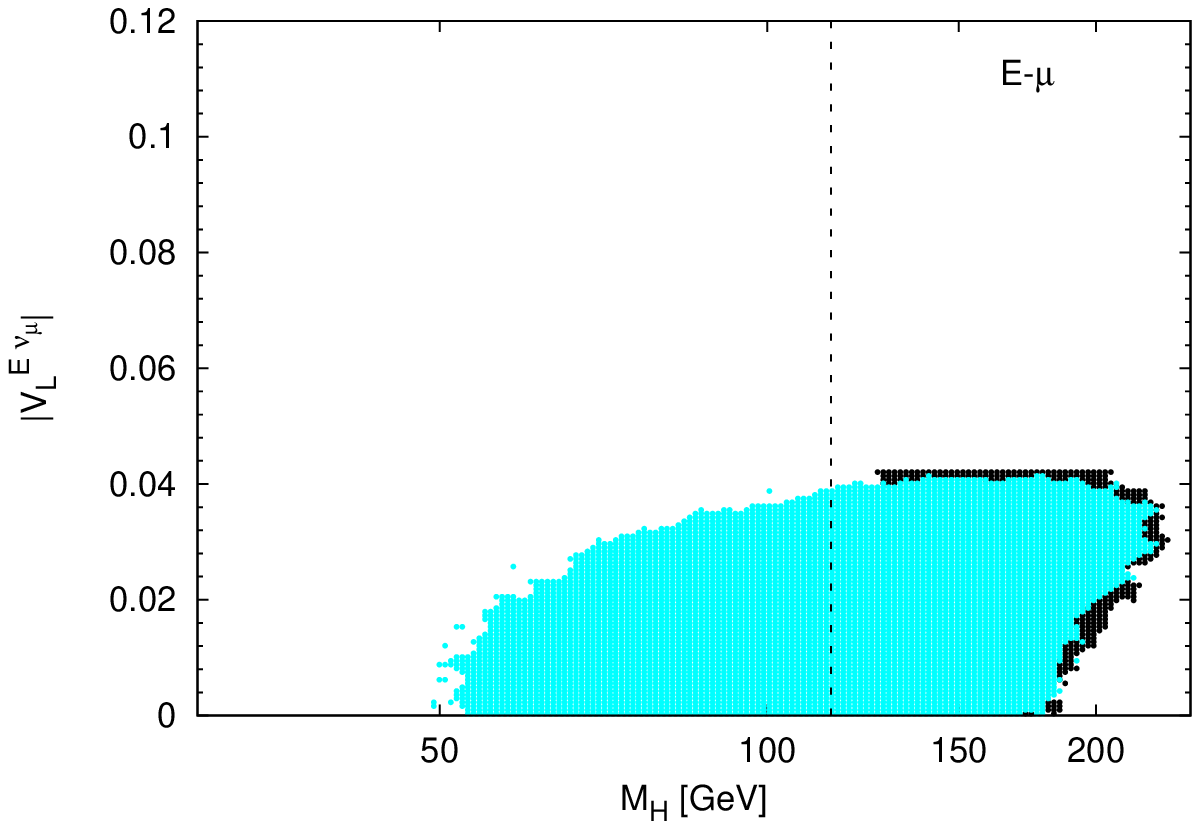}
\includegraphics[width=7cm]{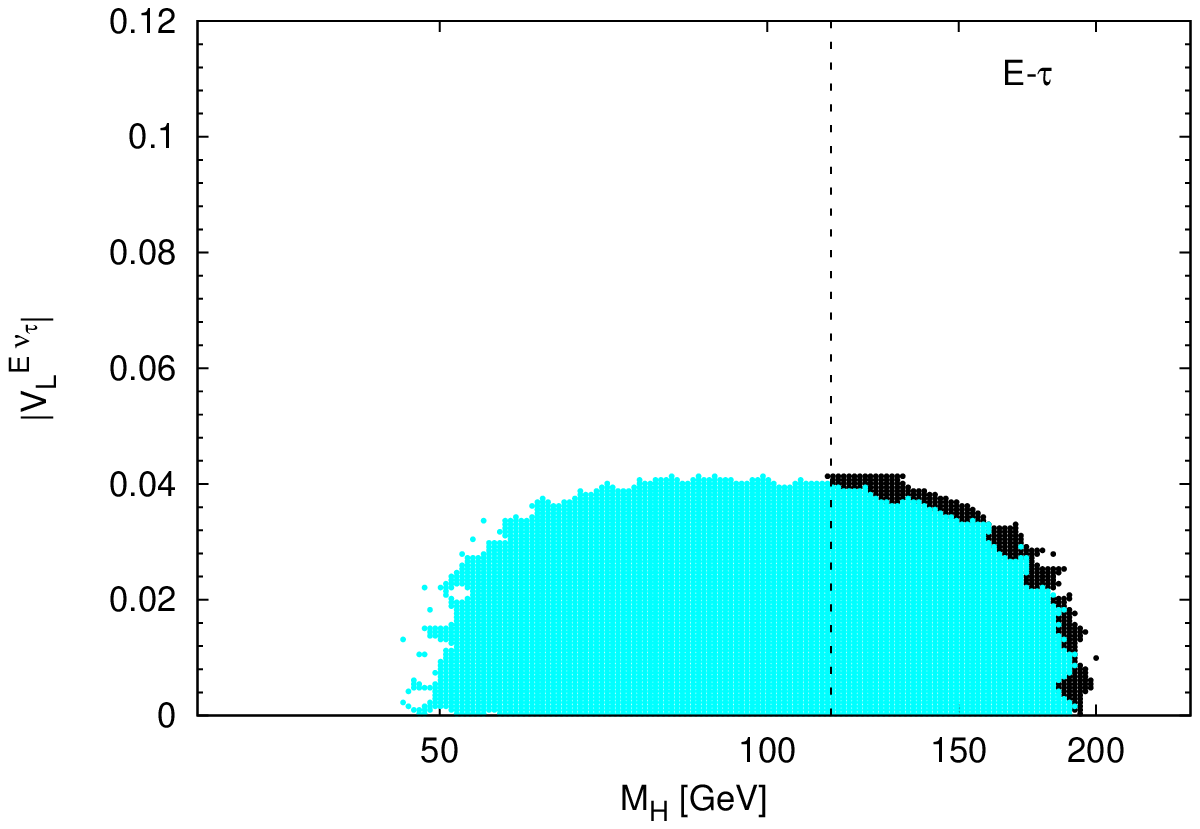}
\includegraphics[width=7cm]{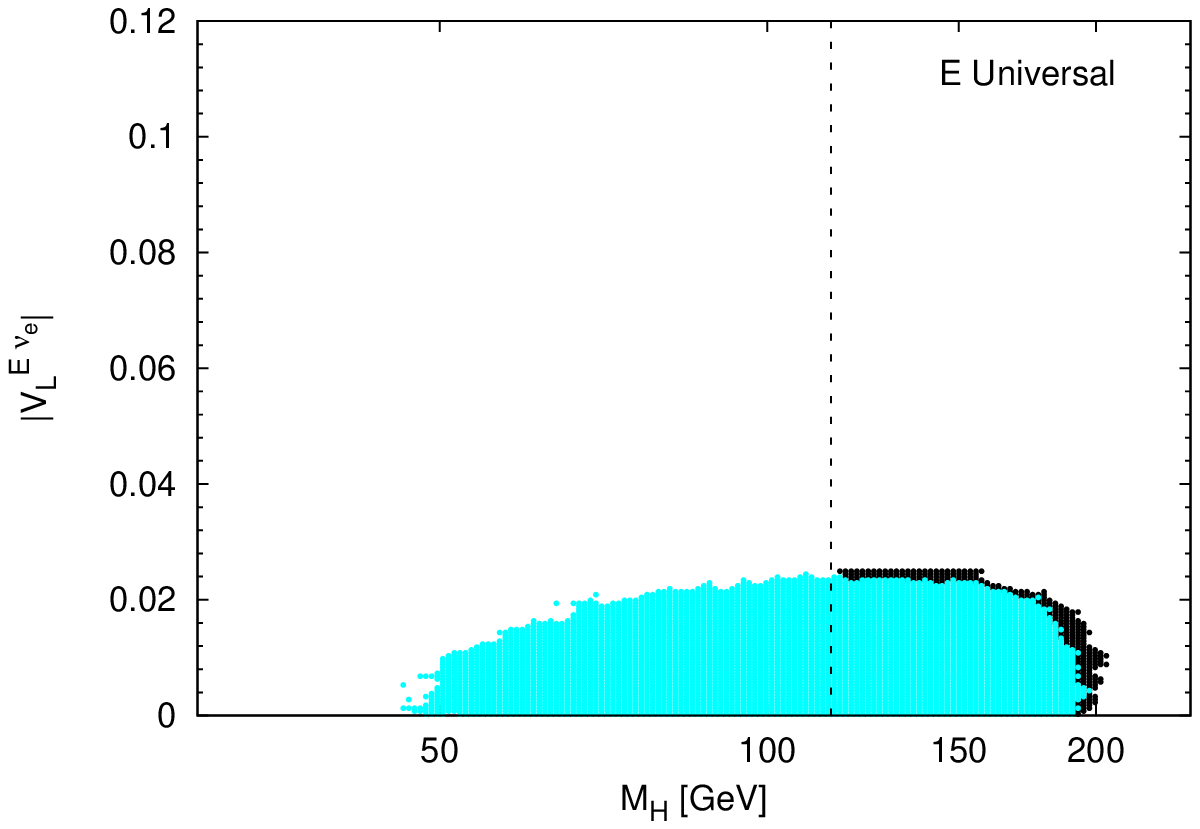}
\caption{90$\%$ confidence region in the $\left|V_{L}^{E \nu}\right|-M_H$
parameter space for the $E$ singlet coupled to the first,
second and third family, respectively. The last plot corresponds
to the universal case. In all cases, the extension of the 90$\%$
confidence region
with the cut $M_H\geq 114.4 \mbox{ GeV}$ (represented by
the vertical dashed line) is shown in black.
\label{LimitplotsE}}  
\end{figure}
%
%
\begin{figure}[t]
\includegraphics[width=7cm]{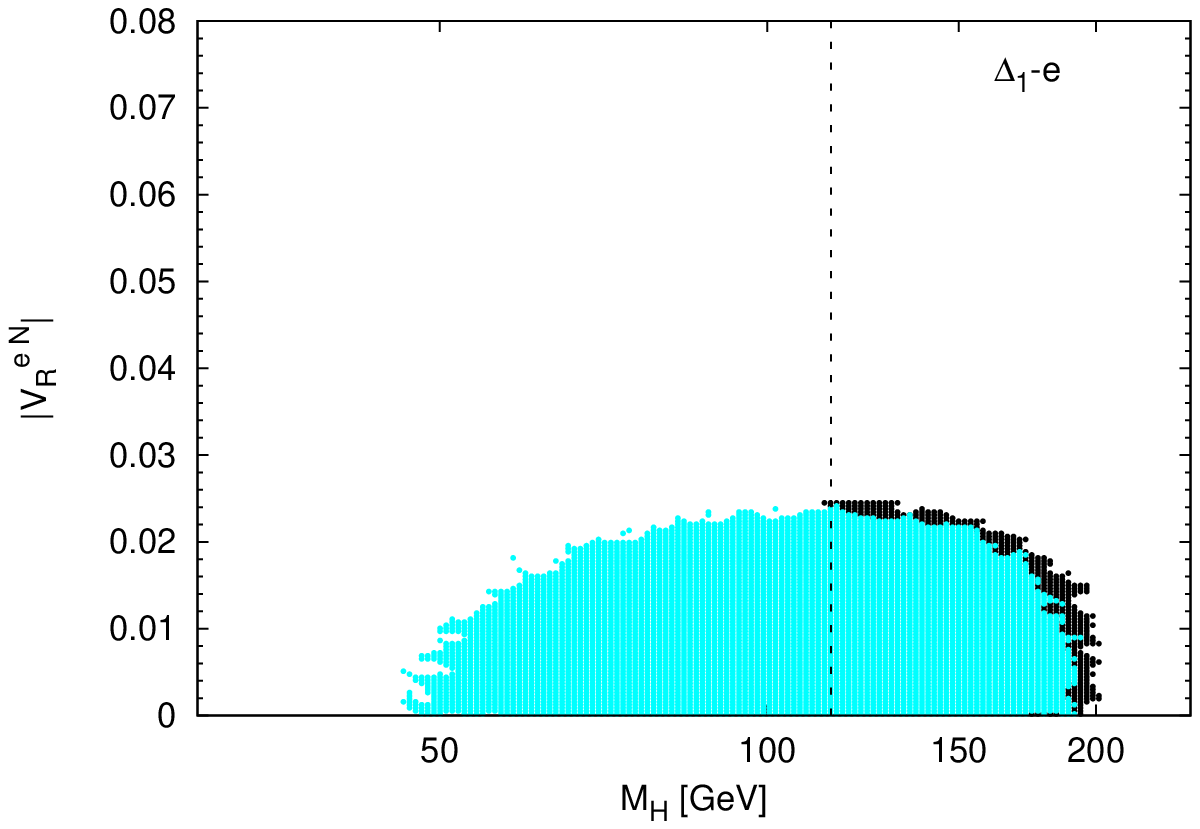}
\includegraphics[width=7cm]{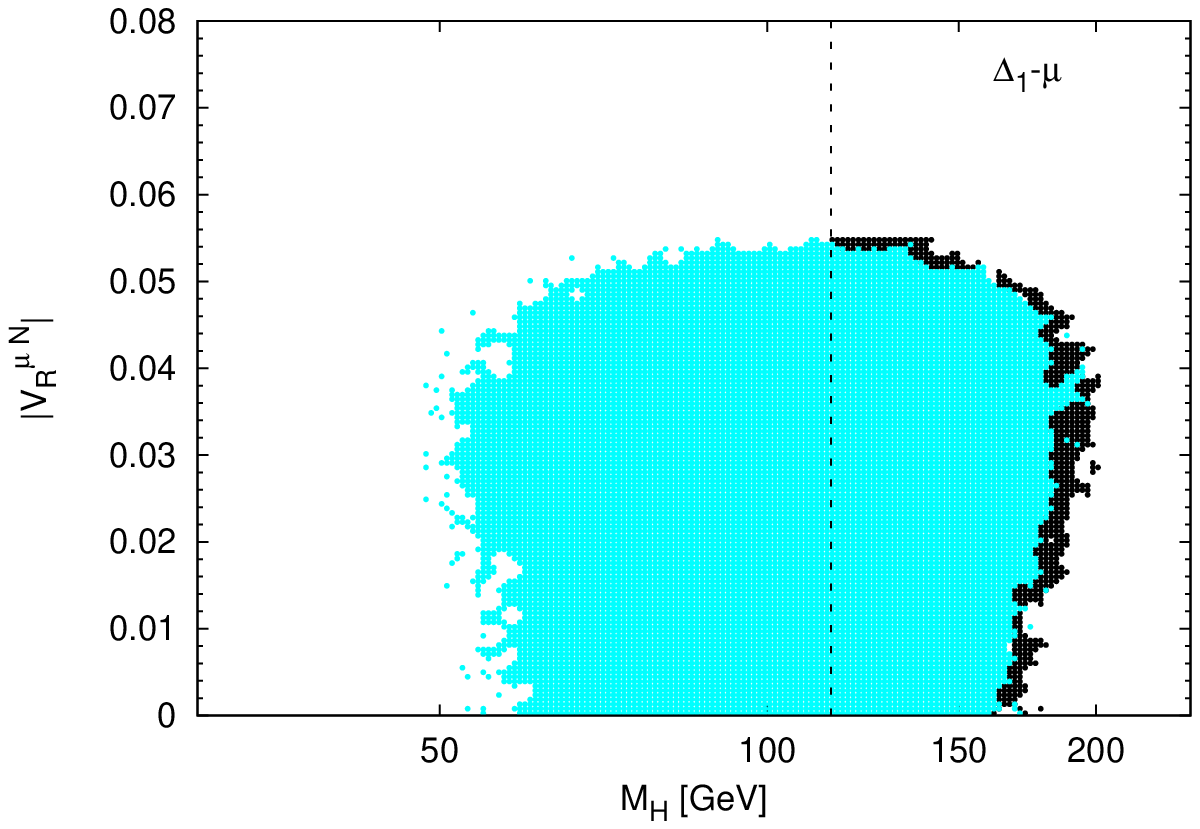}
\includegraphics[width=7cm]{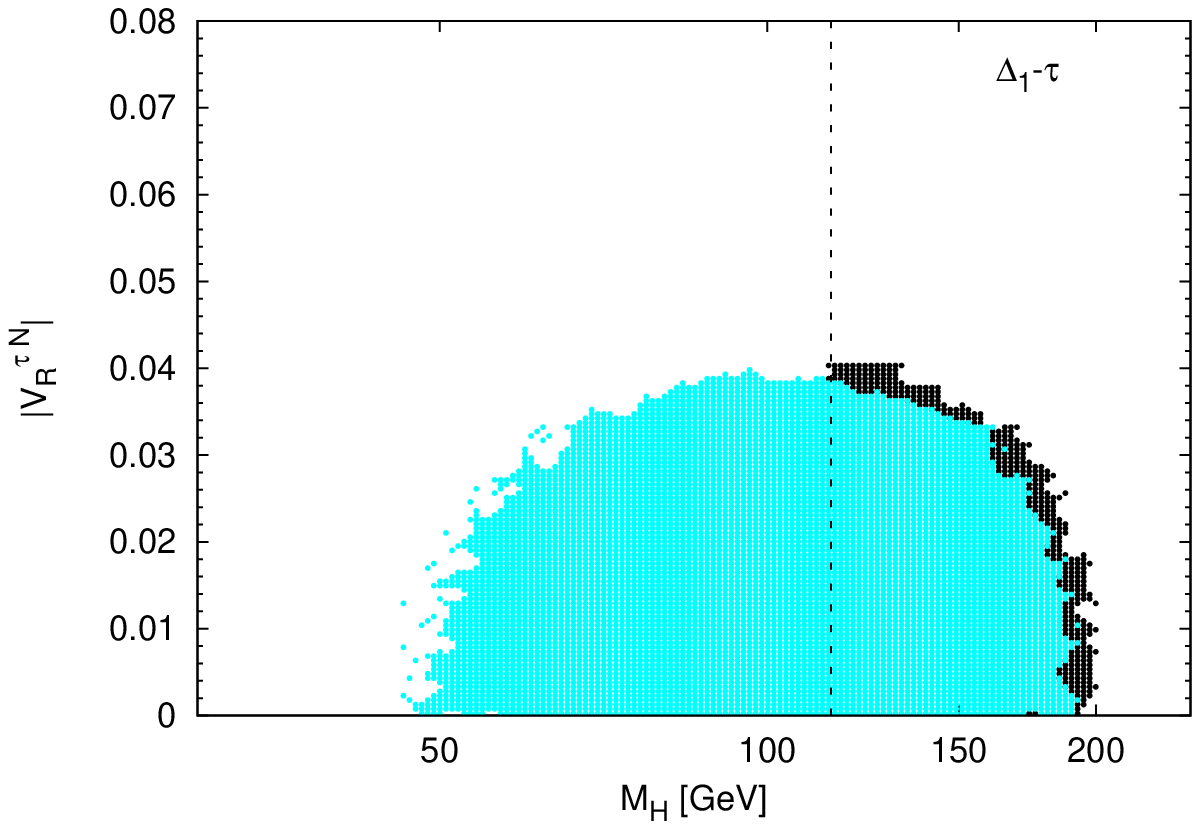}
\includegraphics[width=7cm]{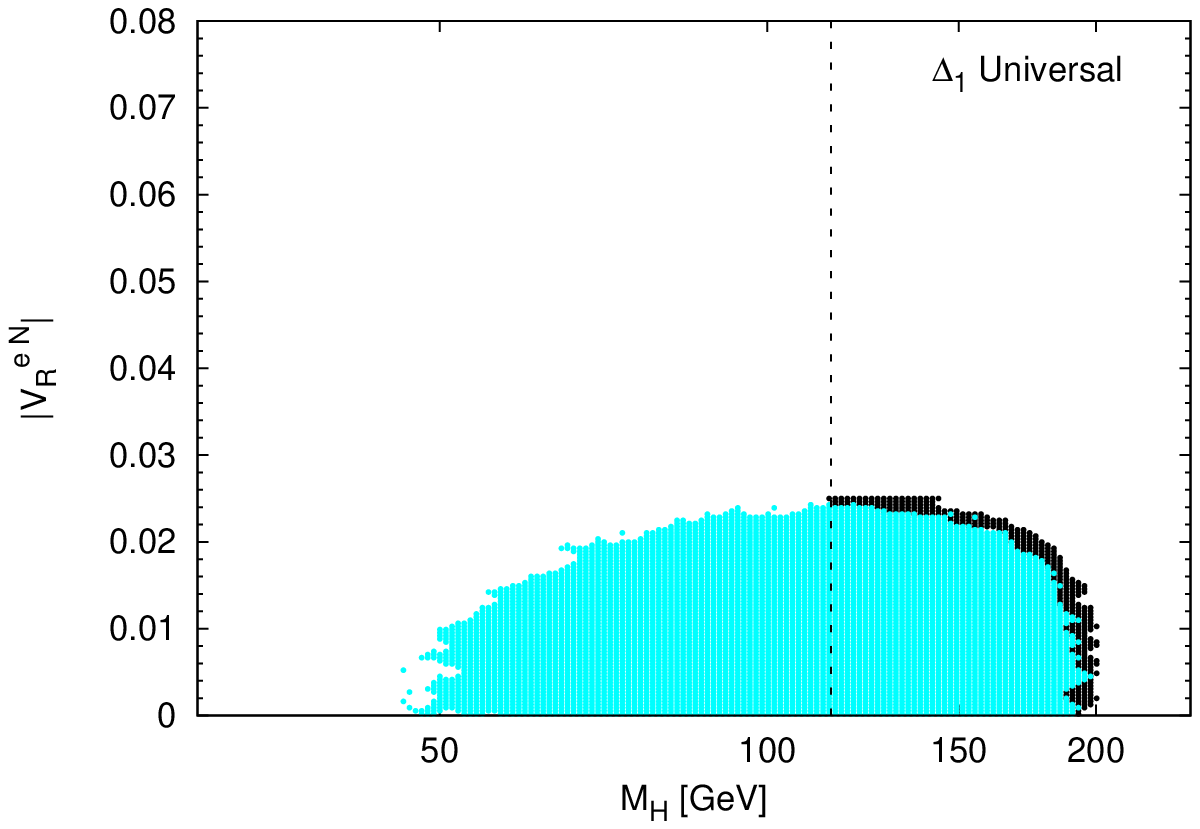}
\caption{90$\%$ confidence region in the $\left|V_R^{eN}\right|-M_H$
parameter space for the $\Delta_1$ doublet coupled to the first, second and third family, respectively. The last plot
corresponds to the universal case. In all cases, the extension of the
90$\%$ confidence region with the cut $M_H\geq 114.4 \mbox{ GeV}$ (represented by
the vertical dashed line) is shown
in black.
\label{LimitplotsNE}}  
\end{figure}
%
%
%
\begin{figure}[t]
\includegraphics[width=7cm]{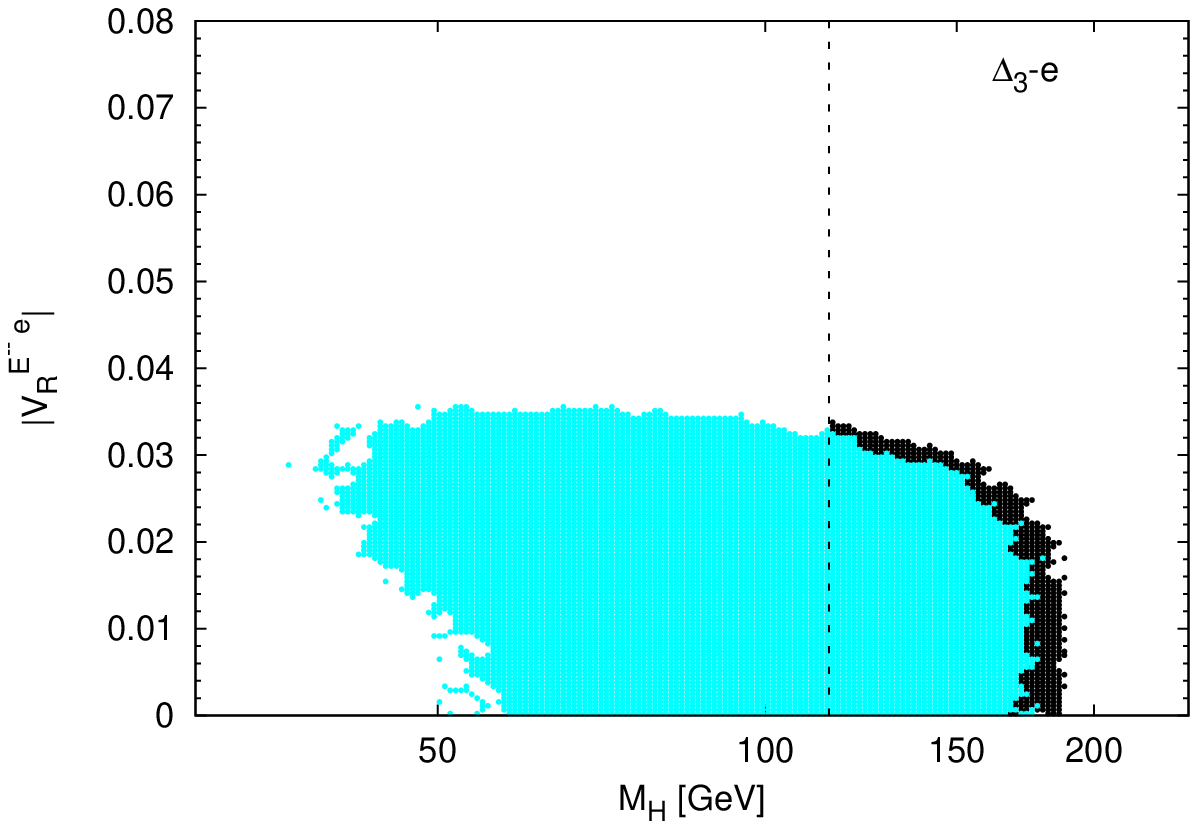}
\includegraphics[width=7cm]{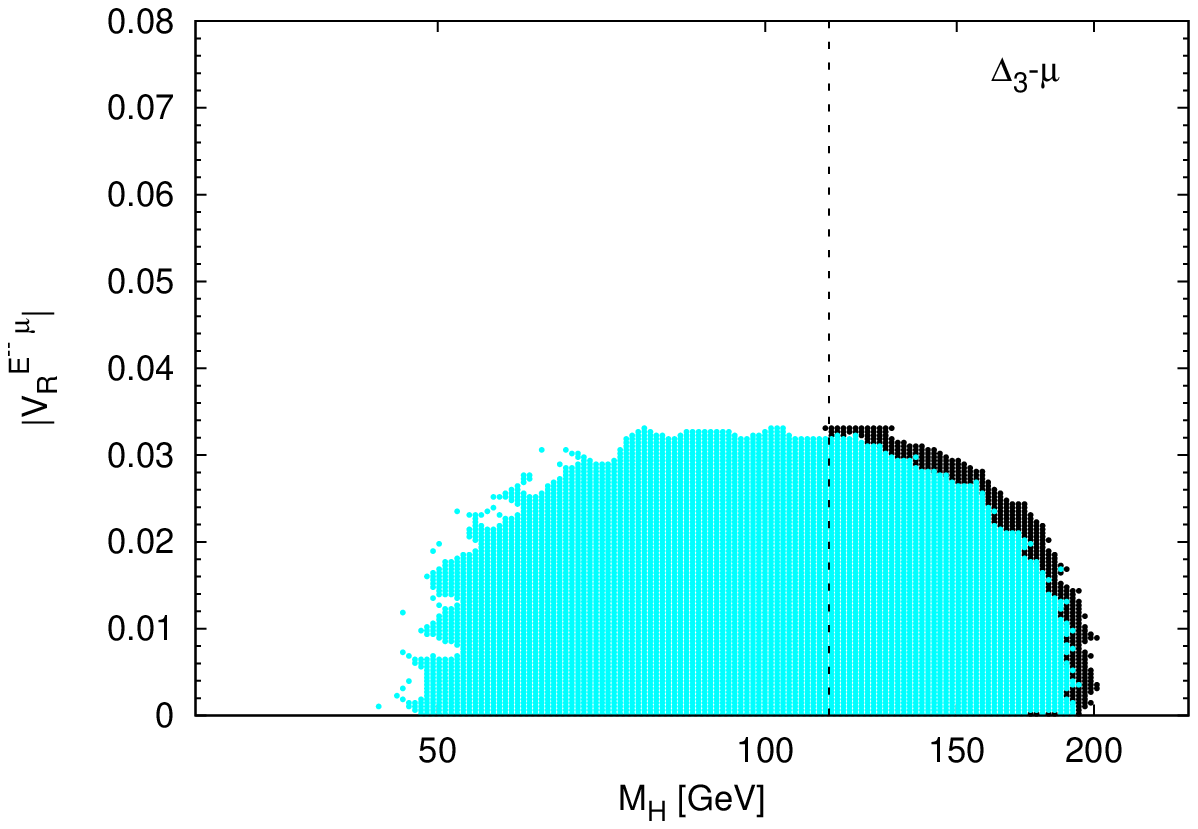}
\includegraphics[width=7cm]{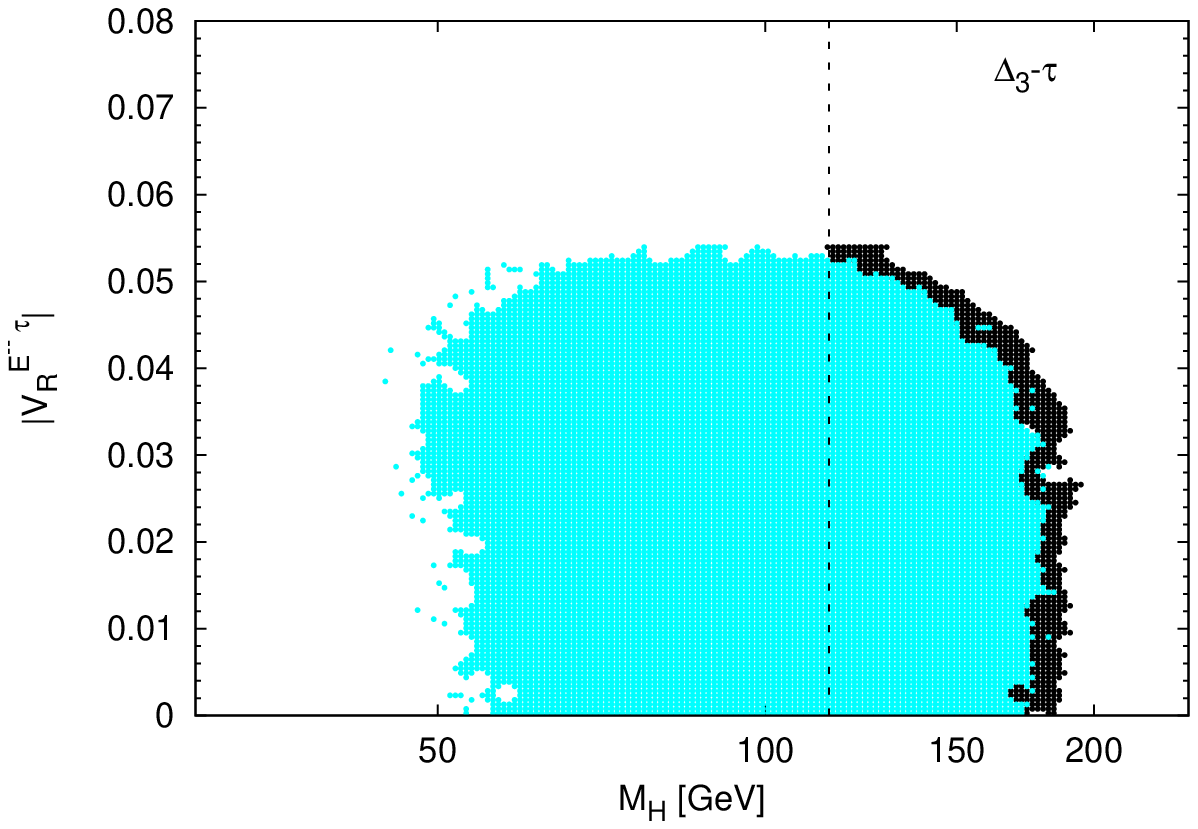}
\includegraphics[width=7cm]{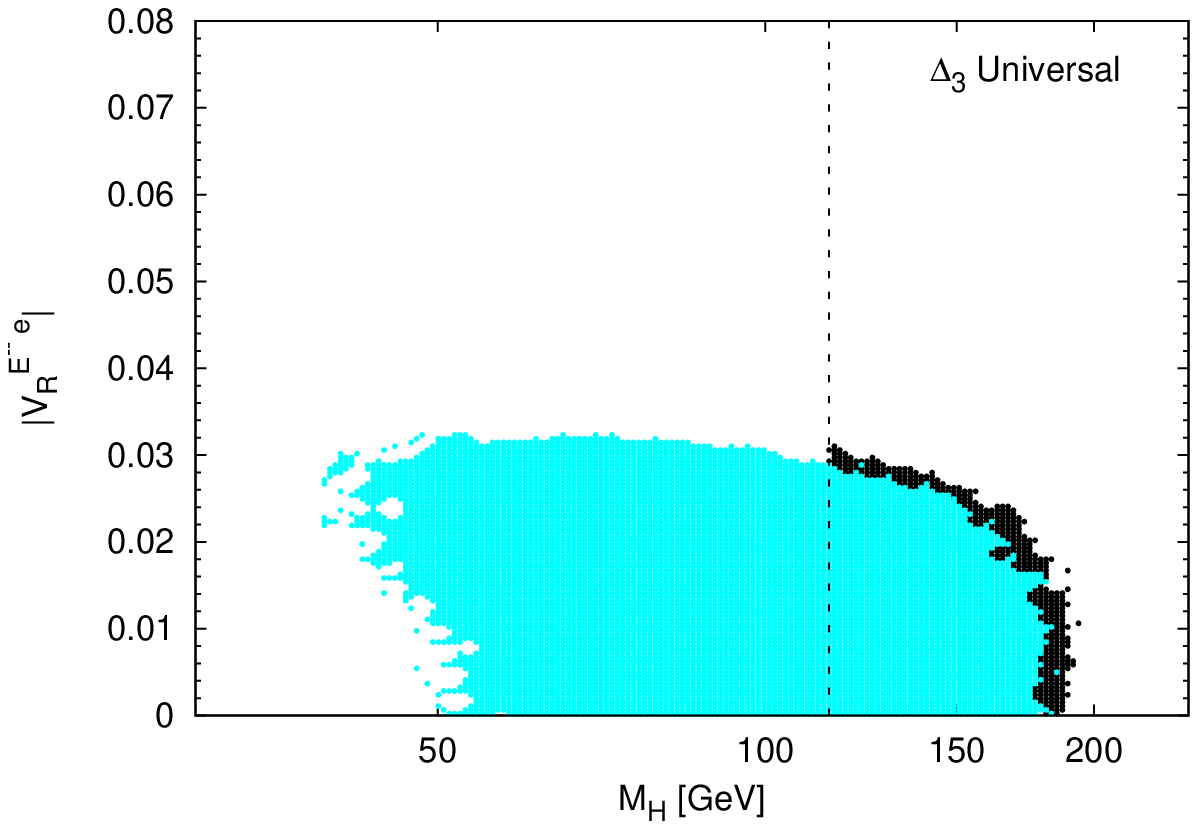}
\caption{90$\%$ confidence region in the $\left|V_R^{E^{--} e}\right|-M_H$
parameter space for the $\Delta_3$ doublet coupled to the first, second and third family, respectively.  The last plot
corresponds to the universal case. In all cases, the extension the
90$\%$ confidence region with the cut $M_H\geq 114.4 \mbox{ GeV}$ (represented by
the vertical dashed line) is shown in black.
\label{LimitplotsEI}}
\end{figure}
%
%
\begin{figure}[t]
\includegraphics[width=7cm]{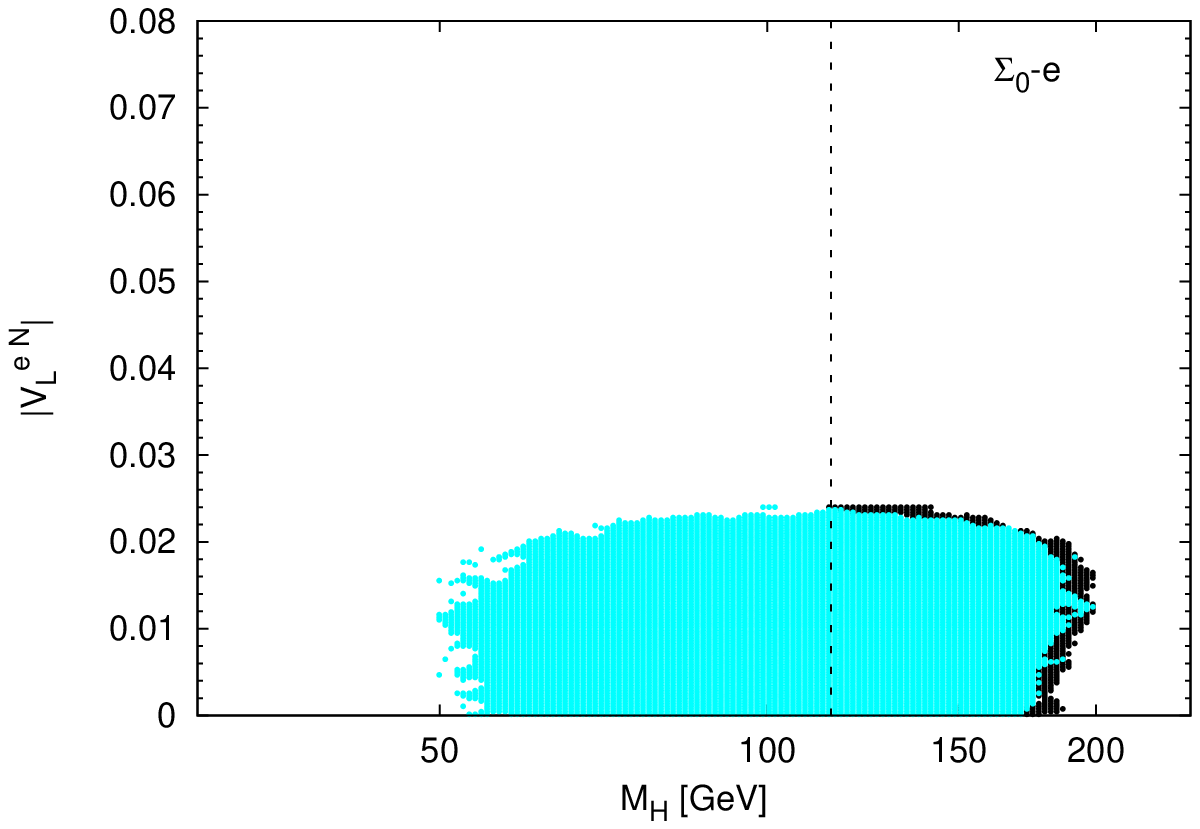}
\includegraphics[width=7cm]{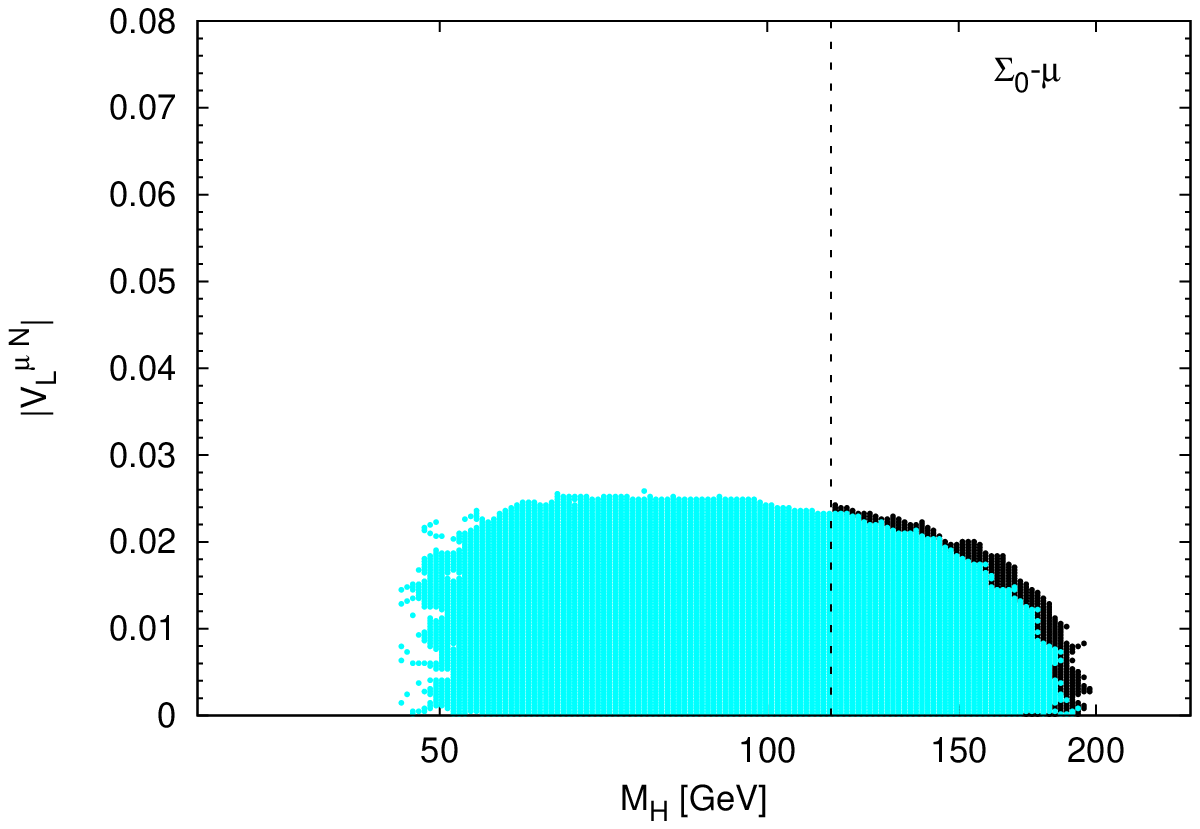}
\includegraphics[width=7cm]{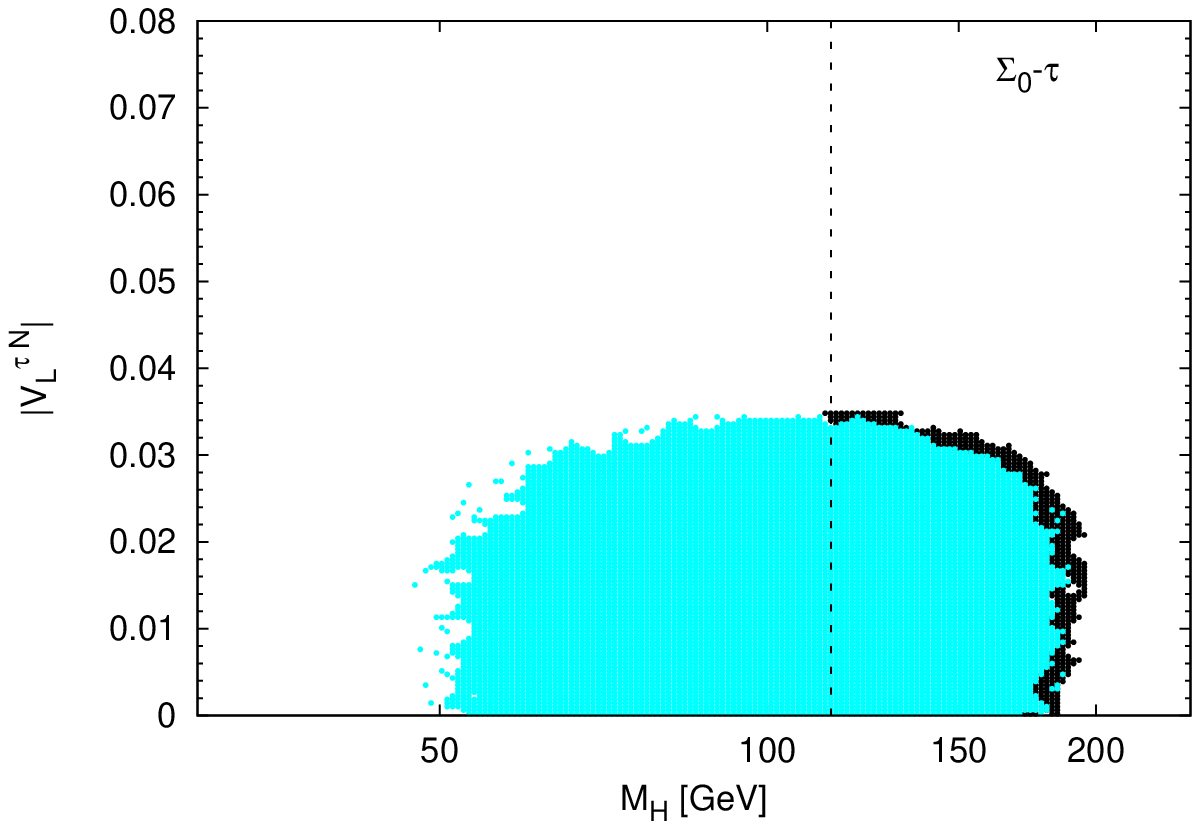}
\includegraphics[width=7cm]{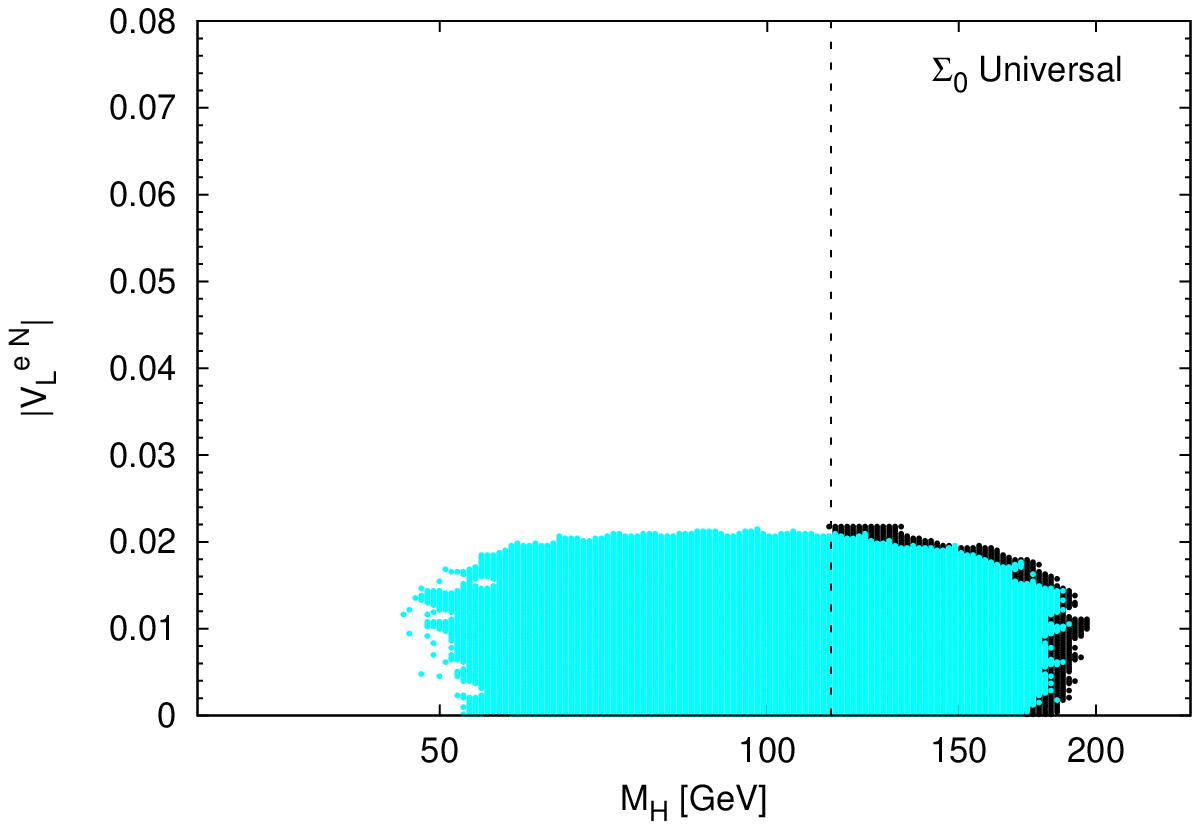}
\caption{90$\%$ confidence region in the $\left|V_L^{e N}\right|-M_H$
parameter space for the $\Sigma_0$ triplet coupled to the first,
second and third family, respectively.  The last plot 
corresponds to the universal case. In all cases the 90$\%$ confidence
region with the cut $M_H\geq 114.4 \mbox{ GeV}$ (represented by
the vertical dashed line) is shown in black.
\label{LimitplotsMNE}}  
\end{figure} 
%
%
%
\begin{figure}[t]
\includegraphics[width=7cm]{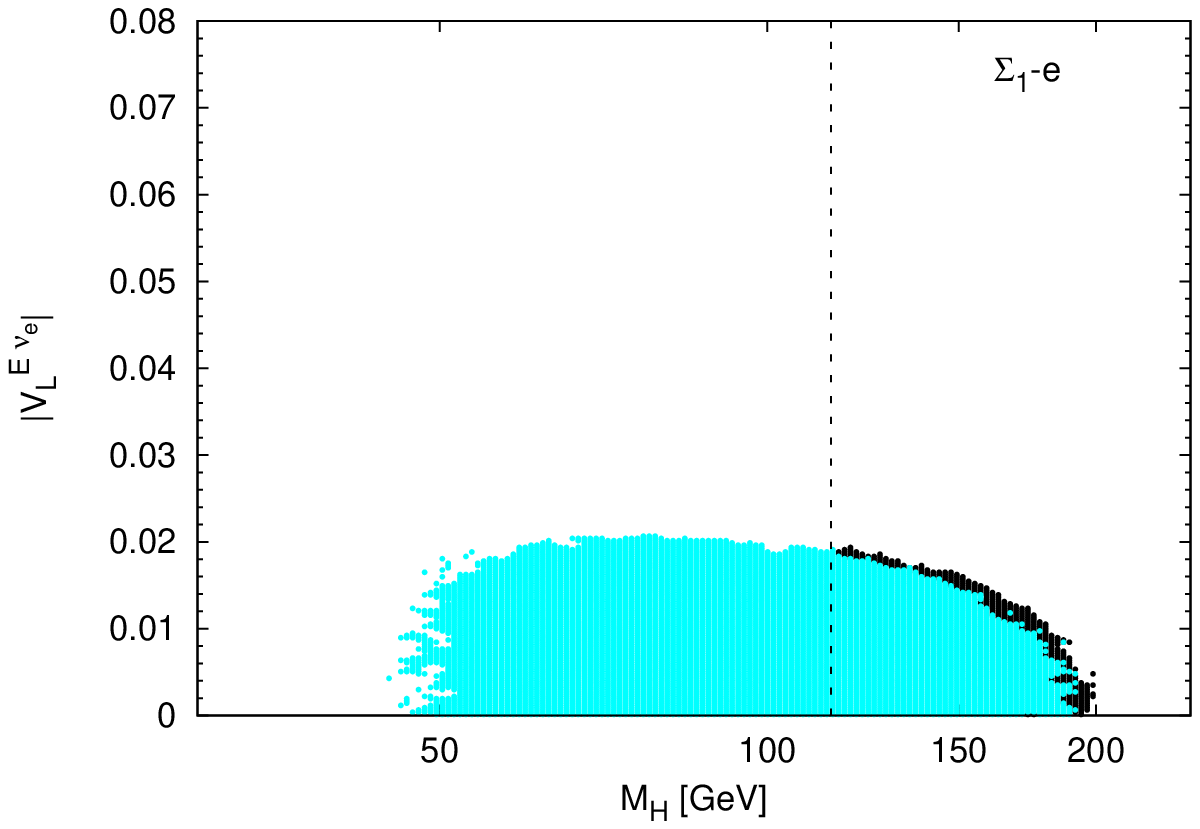}
\includegraphics[width=7cm]{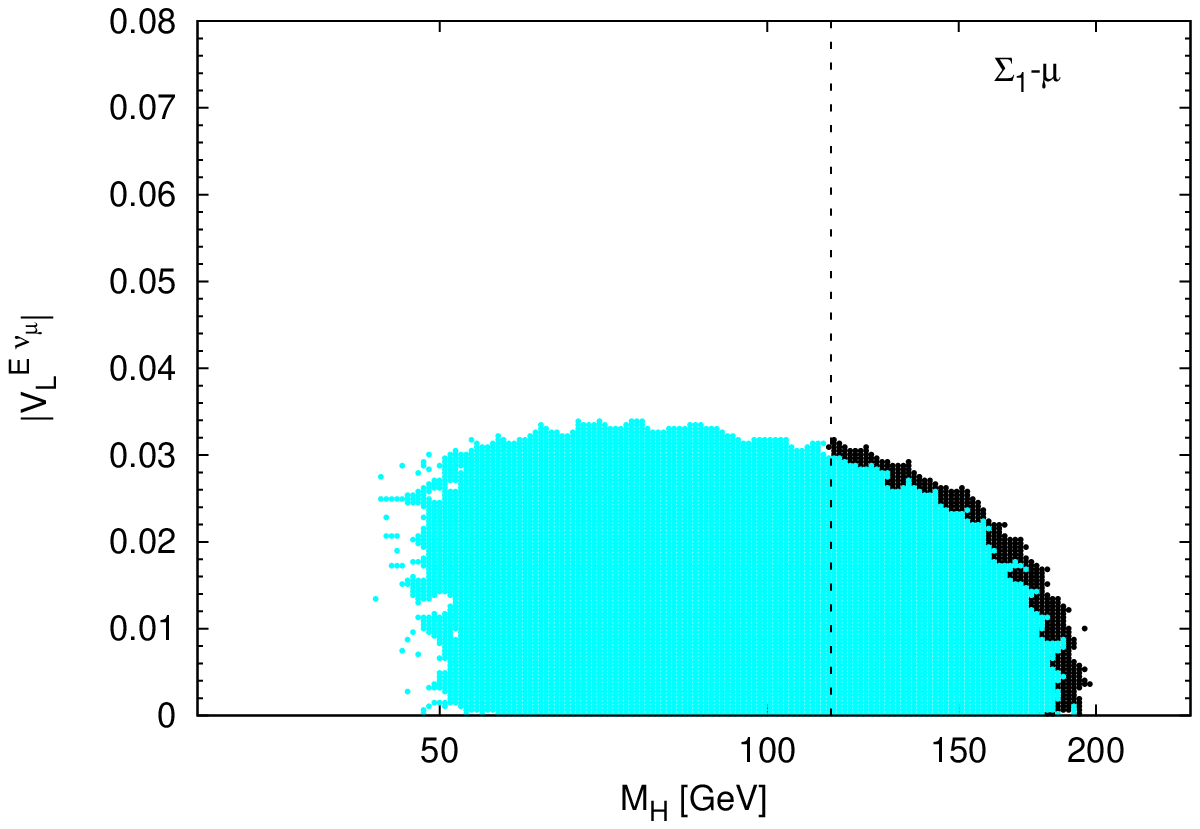}
\includegraphics[width=7cm]{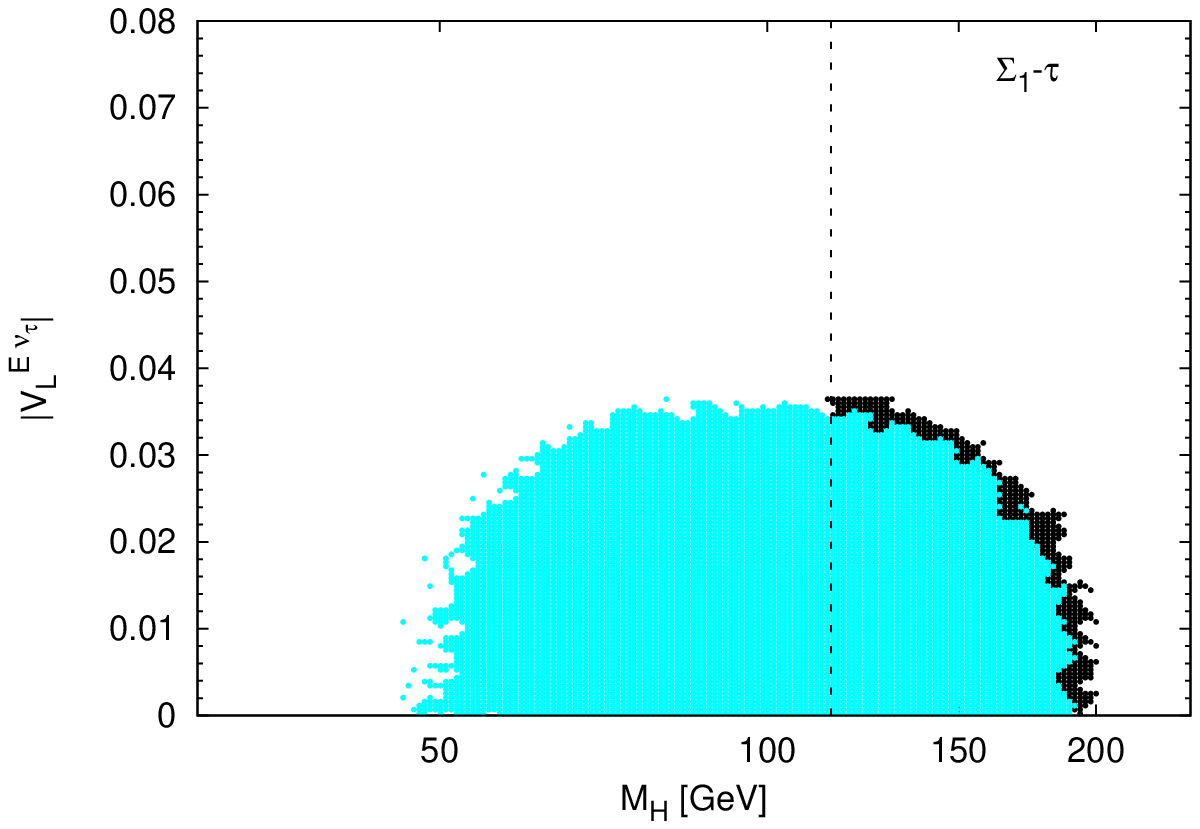}
\includegraphics[width=7cm]{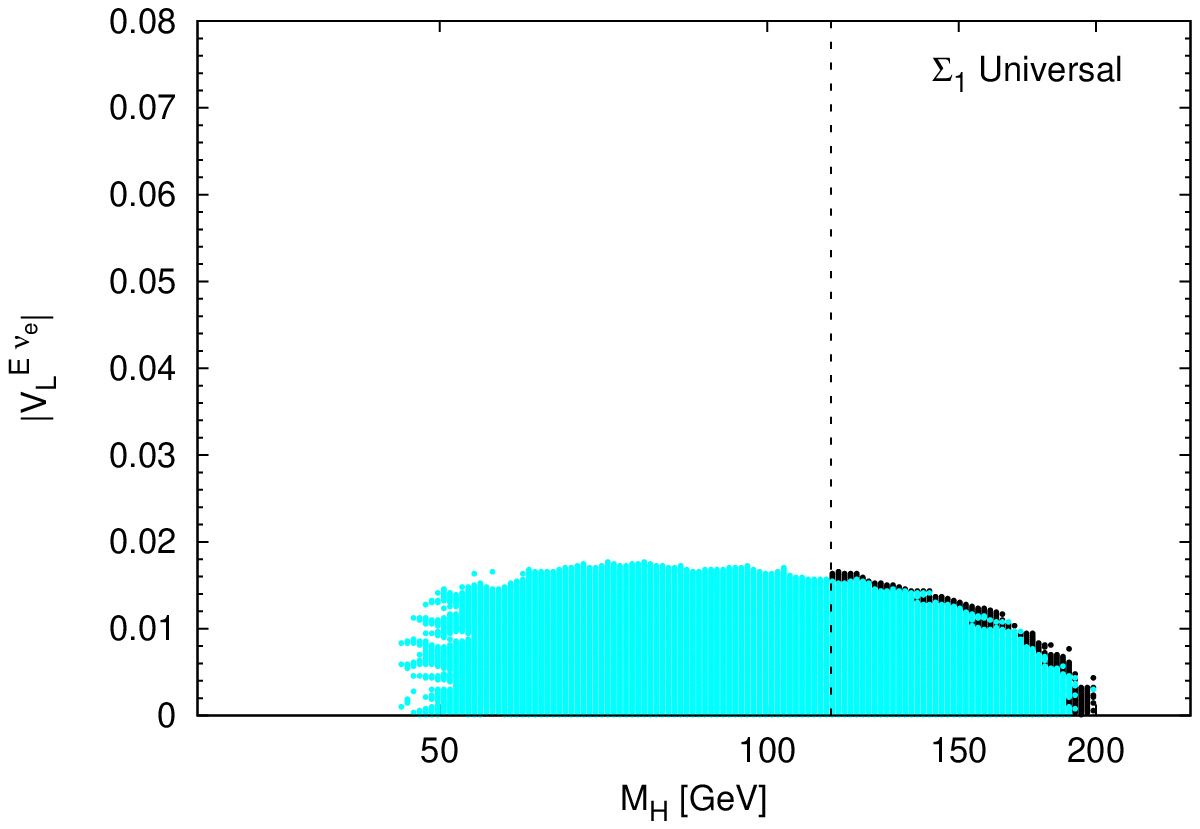}
\caption{90$\%$ confidence region in the $\left|V_L^{E \nu}\right|-M_H$
parameter space for the $\Sigma_1$ triplet coupled to the first,
second and third family, respectively.  The last plot
corresponds to the universal case.  In all cases the 90$\%$ confidence
region with the cut $M_H\geq 114.4 \mbox{ GeV}$ (represented by
the vertical dashed line) is shown in black.
\label{LimitplotsNEI}} 
\end{figure}
%

In Figs~\ref{LimitplotsN} to~\ref{LimitplotsNEI} we show the 90$\%$
C.L. regions in 
the $\left|V\right|-M_H$ parameter space. In these plots we display in
blue the 90\% probability region of the fit without any 
restriction on 
$M_H$, and in black the extension of the 90\% region when we enforce
$M_H\geq 114.4\, \mathrm{GeV}$. The direct lower limit on $M_H$ is
represented by the vertical line.  

As is aparent in the plots, in some cases there is a correlation
between the mixing and $M_H$. In particular, we can see in
Fig.~\ref{LimitplotsN} a strong positive correlation for the singlet $N$,
as long as it mixes with the first and/or second family of SM 
leptons. As a result, the preferred
Higgs mass is larger than in the SM\footnote{This effect has been
discussed before by Loinaz et al.\ in~\cite{Loinaz}. In that
reference, a much 
heavier Higgs is allowed because the constraint from $M_W$ is not
enforced (or it is compensated by unknown new physics). We discuss the
differences between our analysis and the one in~\cite{Loinaz}
below.}. This is in fact responsible for
part of the improvement in the $\chi^2$ in this case.
We analyze the interplay between the
Higgs mass and the mixing of neutrino singlets
in more detail in the next section. In Table~\ref{MHLimits} we give
the $90\%$ C.L. upper limits that we find in
the different scenarios. These limits take
into account the direct lower bound. The limits with extra
singlets are significantly weaker than in the SM.

Because $A_{FB}^{0,b}$ and $g_L^2$ show discrepancies beyond
2.6~$\sigma$, in the SM and in all the extensions with
leptons---except for $(\Delta_3)_e$, which gives a slightly smaller
pull of 2.4 for $A_{FB}^{0,b}$---it is reasonable to consider them as
outliers that should be removed from the fits. This is indeed the
correct approach if the   
anomalies are due to underestimated systematic errors or to (additional)
new physics which does not modify other observables. In the SM,
this would make the preferred Higgs mass much lower than the direct
LEP limit~\cite{Chanowitz:2002cd}. To quantify to what extent this effect is
problematic nowadays and see whether the situation is improved by
extra leptonic singlets, we have repeated the 
fits for the SM and for universal neutrino singlets excluding all
low energy observables and $A_{FB}^{0,b}$, and imposing again the
constraint $M_H\geq 114.4\, \mathrm{GeV}$. We find
$\chi^2/\mathrm{d.o.f.} = 11.1/13$ in the SM and
$\chi^2/\mathrm{d.o.f.} = 7.7/12$ for extra singlets. Therefore,
we see that there is a significant improvement in the quality of this
fit when we include new singlets. This comes in part from the fact that a
bigger improvement in $A_l(\mathrm{SLD})$ is possible when
$A_{FB}^{0,b}$ is disregarded. On the other hand, it is also apparent
that the SM is  
perfectly consistent with this reduced set of data, even with the
constraint from the direct Higgs searches, with a probability of
60.2\,\% of a larger $\chi^2$.

In fact, in general the SM ``adapts'' to relatively large values of
$M_H$ by lowering and increasing a bit the values of $\Delta
\alpha_\mathrm{had}^{(5)}$ and $m_t$, respectively. This is not
necessary with extra neutrino singlets coupled to the first or second
families. In this regard, let us note that $g_\mu-2$ prefers higher
values of $\Delta \alpha_\mathrm{had}^{(5)}$, so that including it in
the fits would favour the extension with singlets with respect to
the SM~\cite{massimo}.

\section{Large neutrino mixing and the Higgs mass}
\label{section_singlets}
From Table~\ref{LeptLimits}, we see that the less constrained extra leptons are
the neutrino singlets. These fields can play the role of see-saw
messengers, although as we have mentioned their contribution to
$\alpha_5$ must be suppressed or cancelled by another contribution. In this
section we analize this case in detail, emphasizing the role
of the Higgs boson.

The mixing of new leptons with the light neutrinos
modifies the invisible width of the Z,
$\Gamma_{\mathrm{inv}}$. This shifts the prediction for
$\sigma_H^0$ in the opposite direction, since
\begin{equation}
\sigma_H^0=12\pi\frac{\Gamma_e \Gamma_h}{M_Z^2\Gamma_Z^2},
\label{SigmaH}
\end{equation}
and $\Gamma_Z= \Gamma_l + \Gamma_h + \Gamma_\mathrm{inv}$ (with the
leptonic width $\Gamma_l=3\Gamma_e$ in the universal case).
For the singlets $N$, the invisible width is smaller and the shift in
$\sigma_H^0$ is positive, so the pull in this quantity is
reduced. These are the only 
effects on Z-pole observables when the new singlet mixes only with the
third family. On the other hand, the independence of these couplings
for different families is limited
in the fit by the decays of the $W^\pm$,
which do not allow for big  
departures from universality in the neutrino couplings. For this
reason, the pull decreases only from $1.7$ in the SM to $0.8$.

A more interesting feature appears as the result of the coupling of
$N$ to the first two families. These couplings generate the
operators $\left(\mathcal{O}^{(3)}_{\phi 
  l}\right)_{ee,\mu\mu}$, which contribute to muon decay and affect
the extraction of the Fermi constant $G_F$ from the muon
lifetime. Because
$G_F$ is an input parameter, this effect
propagates to all observables, giving indirect  
corrections that mimic the ones of the
$T$ oblique parameter of Peskin and Takeuchi~\cite{Peskin:1991sw}.
With the normalization 
of~\cite{Barbieri:2004qk},  
\begin{equation}
\hat{T}_\mathrm{eff}= -\mathrm{Re}\left[\left(\alpha_{\phi
    L}^{(3)}\right)_{ee}+\left(\alpha_{\phi
    L}^{(3)}\right)_{\mu\mu}\right]\frac{v^2}{\Lambda^2} \,.
\label{teff}
\end{equation}
This equation applies to all our observables except $M_W$, which is
discussed below.

As the dominant effects of the Higgs boson are oblique
as well, some cancellations can take place. Indeed, including the leading
contribution of the Higgs mass and the shift in $G_F$, the corrections
to the oblique parameter $\epsilon_1$~\cite{Altarelli:1997et} are
given by
\beq
\delta \epsilon_1=-\frac
 {3G_{F}M_W^2}{4\sqrt{2}\pi^2}\tan^2{\theta_W} \log{\frac{M_H}{M_Z}} +
 \hat{T}_\mathrm{eff}.    
\label{deltaeps}
\eeq
Hence, we see that the effect in $\epsilon_1$ of a heavy Higgs mass
can be compensated by a positive value of
$\hat{T}_\mathrm{eff}$. In fact, it is known that a heavy Higgs can be
made consistent with EWPD by new oblique physics that gives a
positive $T$ parameter, even if the positive contributions of the Higgs
to $\epsilon_3$ are not cancelled by a negative $S$ parameter.
For the neutrino singlets, the sign of $\hat{T}_\mathrm{eff}$ is
actually positive. This, combined with the improvement in the
hadronic width, explains that the fit prefers relatively large values
of $M_H$, as can be seen in Tables~\ref{MHLimits} and~\ref{MHFitN}. 
In Fig.~\ref{LimitplotsN} we observe clearly how a non-vanishing
mixing of new 
singlets with electron and/or muon neutrinos allows for larger values of
$M_H$, thus eliminating the (mild) tension between the global
electroweak fit and the direct LEP lower bound on $M_H$.
%
\begin{table}[ht]
\begin{center}
{\scriptsize
\begin{tabular}{| c | c c c c |}\hline
$ $&$ $&$ $&$ $&$ $\\[-.25cm]
$N\mbox{ Coupling}$&$e$&$\mu$&$\tau$&$\mbox{Universal}$\\[.1cm]
\hline\hline
$ $&$ $&$ $&$ $&$ $\\[-.25cm]
$M_H\left[\mbox{GeV}\right]$&
$132.4 $&$135.9 $&$114.4 $&$135.4 $\\[.1cm]
\hline
\end{tabular}}
\caption{Best-fit values of the Higgs mass (in $\mathrm{ GeV}$) for
  the extensions with neutrino singlets. In all the other cases the
  Higgs mass prefers to remain at the imposed cut,
  $M_H=114.4~\mbox{GeV}$.}  
\label{MHFitN}
\end{center}
\end{table}
%

On the other hand, unlike the shift in $G_F$, a genuine $T$ parameter
from new oblique physics would 
give additional direct contributions to $M_W$ (for fixed $M_Z$). These
are not included in our $\hat{T}_\mathrm{eff}$, and in general cannot
be generated by any kind of new leptons at tree level. A heavy Higgs
gives the complete $T$-like contributions (in addition to $S$-like and
suppressed $U$-like contributions). Therefore, there is no cancellation
of Higgs and singlet effects in $M_W$, once the relation
between mixings and $M_H$ has been determined from Z-pole
observables. This prevents the Higgs 
from being too heavy, and the lepton mixings from being too large. 

Let us also note that the net contribution of the new singlets to
neutrino--nucleon deep inelastic scattering is suppressed, due to an
approximate cancellation between their indirect and direct
effecs. Therefore, the dominant effect is the oblique Higgs
boson contribution, which is negative 
when $M_H$ is increased with respect to the reference
value\footnote{Alternatively, the Z-pole observables impose an
  approximate cancellation between the $M_H$ and
  $\hat{T}_\mathrm{eff}$ contributions. This leaves the negative
  direct contribution of the new singlets.}. 
This would explain the NuTeV anomaly if the Higgs were allowed to
be very heavy. But as we have discussed above, $M_W$ prefers a light
Higgs, and in the best fit to all observables there is no improvement in
$g_L^2$.

Our conclusions are not at odds with the one of
Loinaz {\it et al.}\ in~\cite{Loinaz}\footnote{As a technical point, let us
  mention that our formulas for $g_L^2$ and $g_R^2$ in neutrino deep
  inelastic scattering differ from the ones in this reference, because
  we include the heavy-lepton contributions to the determination of
  $V_{ud}$ from $\beta$ decay, just as we did for $G_F$ and muon decay. These
  contributions reverse the sign of the
  singlet contributions to $g_L$, which then play against the improvement
  of the NuTeV anomaly. However, in both cases the singlet
  contributions are subleading with respect to the Higgs ones and do
  not alter the qualitative conclusions.}. They claim that mixing of light and
heavy neutrinos can account for the NuTeV anomaly and, together with a
heavy Higgs, give an excelent fit {\em as long as} $M_W$ is not included
in the fit or additional new physics supplies a big $U$
parameter. We have preferred, instead, to include $M_W$ in
our fits, as this observable is well measured nowadays. Moreover,
dimensional and symmetry arguments suggest that, in 
the absence of fine tuning, $U$ 
is smaller than $T$ for any new physics coming in at a
scale larger than $M_W$~\cite{Peskin:1991sw,Barbieri:2004qk}. This is indeed
found in known calculable models. So, it seems difficult that any new
physics can yield the values $U \gg T$ required in the fit
of~\cite{Loinaz}\footnote{We do not claim that this possibility is logically
excluded. The authors of~\cite{Loinaz} propose the possibility that threshold
effects in a strongly coupled sector might give rise to the necessary
enhancement of $U$.}.
When $M_W$ is included in the global fit and no {\it ad
  hoc\/} $U$ parameter is introduced to eliminate its influence, the
results are not that spectacular. We find that the Higgs cannot be
very heavy and that the NuTeV anomaly is not explained. Nevertheless,
as we have discussed, there is an improvement in $\sigma_H^0$ (through
the invisible width) and a Higgs heavier than in the SM is allowed.


\section{Conclusions}

We have performed a global fit to existing EWPD for extensions of the
SM with new vector-like leptons. The analysis makes use of the corresponding 
effective Lagrangian up to dimension 6, which is justified by the smallness of 
the mixings we find. 
Our main results are displayed in Tables~\ref{LeptLimits} and
\ref{MHLimits}, and 
illustrated in the different plots. 
In the cases that had been analyzed
before~\cite{Langacker:1988ur,Nardi:1994iv,Bergmann:1998rg}, we 
find more stringent limits (at the few per cent level). This reflects
the better agreement of the SM predictions with the present data.

In Table~\ref{LeptFits}, we give the
improvements in the $\chi^2$ of the global fit when the SM is
supplemented by new leptons. 
The addition of more than one type of extra lepton multiplet at a time
does not improve the quality of the fit. The $\chi^2/$d.o.f. is
(slightly) reduced with respect to the one in the SM for $(\Delta_1)_\mu$ only.
Even if we do not find any significant improvement of the SM global fit,
it is interesting to observe that TeV-scale vector-like leptons with
sizeable mixings are 
consistent with EWPD. An interesting feature of the fits is that the
presence of extra singlets mixing 
with the electron and/or muon neutrinos favours higher values of the
Higgs mass, which lie confortably in the region allowed by direct searches
of the Higgs at LEP. This accounts for part of the improvement in the
$\chi^2$ in these cases, and implies significantly weaker upper bounds
on the Higgs mass. For mixing with muon neutrinos, $M_H <
267\,\mathrm{GeV}$ (90\% C.L.), with the best-fit value $M_H=136\,
\mathrm{GeV}$. Conversely, such extra lepton singlets
would be favoured with respect to the SM if the Higgs were
eventually found to be heavy.
We have also seen that an explanation of the NuTeV
anomaly by the mixing of the SM neutrinos with extra neutrinos is
precluded, in the absence of additional new physics, by the
constraints imposed by other electroweak observables.

In Table~\ref{LeptLimits}, we collect 
the 90$\,\%$ C.L. bounds and the corresponding best values for the
mixings between the different possible heavy vector-like leptons and the  
SM fermions. The mixing with the SM leptons can be as large as  
$|V^{\tau N}_L| \sim$ 0.079 at 90 $\%$ C.L. for heavy neutrino singlets 
mixing only with the third family.
Other mixings are bounded to be less than $\sim$ 0.06 at 90 $\%$ C.L..
They are independent of the Dirac or Majorana character of the new leptons.
 
These limits have consequences for heavy lepton production and decay 
at large colliders.
At LHC, they are in general more efficiently produced in pairs
\cite{delAguila:1989rq},  
except for heavy neutrino singlets, which have to be single produced 
in association with SM leptons through their mixing, as they have
no other SM interactions. 
In this case the new limits $|V^{eN}_L| <$ 0.055 and $|V^{\mu N}_L| <$ 0.057 are
better than those found previously, $|V^{eN}_L| <$ 0.074 and 
$|V^{\mu N}_L| <$ 0.098 \cite{Bergmann:1998rg}. Therefore, the
small parameter space which may be reached at the
LHC~\cite{Han:2006ip} is further reduced.  
For instance, heavy Majorana neutrino singlets coupling only to muons may be
observable at LHC for masses below 200 GeV. This limit can be much 
higher, however, in the presence of other interactions, up to 2
TeV for new  
right-handed gauge bosons of a similar mass and with a standard gauge coupling 
strength \cite{Datta:1992qw} (see for a review \cite{delAguila:2008iz}). 
Dirac neutrino singlets are expected to be beyond the LHC reach. 
All other lepton additions can be pair produced, and then their
discovery limit  
does not depend on the mixings, 
which only enter in the decay rates and are still large enough to
allow the heavy leptons decay inside the detector. Hence, their rough
discovery limit is near the TeV scale~\cite{delAguila:1989rq}. On the
other hand, at $e^+\,e^-$ colliders the main production mechanism is
through mixing with the first family. For instance, a neutrino singlet
mixing with the electron neutrino with $|V_L^{eN}|>0.01$ is allowed by
our bounds and would be observed at ILC
for masses $M_N<400 \, \mathrm{GeV}$, and at CLIC for $M_N < 2\,
\mathrm{TeV}$~\cite{del Aguila:2005pf}. On the other hand, these
stringent limits also makes more difficult the observation of possible
deviations from unitarity in neutrino oscillations~\cite{Bekman:2002zk}.

Vector-like leptons at the TeV scale appear naturally in many models,
for example those with extra dimensions  
or larger gauge symmetries at low energy.
As already emphasized, the new singlets and triplets of zero
hypercharge can be Majorana and act as see-saw messengers of type~I and~III,
respectively. If these fields exist with 
large mixings and relatively light masses, their contributions to
neutrino masses  
and neutrinoless double $\beta$ decay 
must be in general suppressed by extra, almost exact symmetries, typically LN 
\cite{delAguila:2007ap,Kersten:2007vk}. 
Thus, in general new leptons at the TeV scale and with relatively
large mixings  
with the SM fermions must be (quasi)Dirac. 
If they are Majorana, the model must include a very efficient
cancellation mechanism  
with an extended field content highly tuned \cite{Ingelman:1993ve}. 

The theory must also incorporate a rather precise 
alignment of the SM charged leptons and the new mass eigenstate leptons: 
each heavy lepton must mix mostly with only one light charged lepton
to fulfill  
the limits on FCNC
\cite{Tommasini:1995ii,Abada:2007ux,Raidal:2008jk}. The 
corresponding limits are a factor 3 to 60  
times more stringent than the flavour conserving ones, derived here. 
This justifies neglecting FCNC effects in our analysis, but also implies 
a strong constraint on definite models.

Finally, it is interesting to study how our conclusions would change
in the presence of other new particles, which are actually present in
many of the models mentioned above. We expect that the interference will
be constructive in many cases. We have checked, for instance, that the
new leptons can 
further improve the global fit of the extra-quark solution to the
$A_\mathrm{FB}^{0,b}$ anomaly proposed in~\cite{Choudhury:2001hs}. The
effective formalism 
we used here is particularly convenient to perform fits involving many
different kinds of new particles~\cite{In preparation}.

\section*{Aknowledgements}

We thank J.A. Aguilar-Saavedra, F. Cornet, P. Langacker,
M. Passera, J. Santiago and J. Wudka for helpful discussions.   
This work has been supported by MEC project FPA2006-05294 and
Junta de Andaluc{\'\i}a projects FQM 101, FQM 00437 and FQM03048.
J.B. also thanks MEC for an FPU grant.



\begin{table}[p]
\section*{Appendix}
\vspace{.5cm}
\begin{center}
{\scriptsize
\begin{tabular}{|c c||c|c|c|c|c|} \hline
$ $&$ $&$ $&$ $&$ $&$ $&$ $\\[-0.2cm]
Quantity &&Experimental Value& Standard Model&Pull&Extended Model&Pull\\ 
         &&                  &               &    &with $\Delta_1$ coupled to $\mu$&\\\hline \hline
$ $&$ $&$ $&$ $&$ $&$ $&$ $\\[-0.25cm]
$m_t$&$\cite{Tevatronmt}$&$172.6\pm 1.4$&$172.9$&$-0.2$&$172.9$&$-0.2 $\\
$\Delta \alpha^{(5)}_\mathrm{had}\left(M_Z^2\right)$&$\cite{Burkhardt:2005se}$&$0.02758\pm 0.00035$&$0.02755$&$+0.1$&$0.02757$&$0.0 $\\[0.1cm]
$\alpha_S\left(M_Z^2\right)$&$\cite{Yao:2006px}$&$0.1176\pm 0.002$&$0.1181$&$-0.3$&$0.1176 $&$0.0 $\\[0.1cm]\hline
$ $&$ $&$ $&$ $&$ $&$ $&$ $\\[-0.25cm]
$M_W\left[\mbox{GeV}\right]$&$\cite{Grunewald:2007xt}$&$80.398\pm 0.025$&$80.365$&$+1.3$&$80.365 $&$+1.3 $\\
$\mbox{Br}\left(W\rightarrow e\nu\right)$&$\cite{Yao:2006px}$&$0.1075\pm0.0013$&$0.1083$&$-0.6$&$0.1083 $&$-0.6 $\\
$\mbox{Br}\left(W\rightarrow \mu\nu\right)$&$ $&$0.1057\pm0.0015 $&$ $&$-1.7$&$ $&$-1.7 $\\
$\mbox{Br}\left(W\rightarrow \tau\nu\right)$&$ $&$0.1125\pm0.0020 $&$ $&$+2.1$&$ $&$+2.1 $\\[0.1cm]\hline
$ $&$ $&$ $&$ $&$ $&$ $&$ $\\[-0.25cm]
$M_Z\left[\mbox{GeV}\right]$&$\cite{LEPEWWG06}$&$91.1876\pm 0.0021$&$91.1876$&$0.0$&$91.1875 $&$0.0 $\\
$\Gamma_Z \left[\mbox{GeV}\right]$&$ $&$2.4952\pm 0.0023$&$2.4952$&$0.0$&$2.4947 $&$+0.2 $\\
$\sigma_H^0\left[\mbox{nb}\right]$&$ $&$41.541\pm 0.037$&$41.480$&$+1.7$&$41.489 $&$+1.4 $\\
$R^0_e$&$ $&$20.804\pm 0.050$&$20.739$&$+1.3$&$20.735 $&$+1.4 $\\
$R^0_\mu$&$ $&$20.785\pm 0.033$&$20.739$&$+1.4$&$20.781 $&$+0.1 $\\
$R^0_\tau$&$ $&$20.764\pm 0.045$&$20.786$&$-0.5$&$20.782 $&$-0.4 $\\
$A^{0,e}_\mathrm{FB}$&$ $&$0.0145\pm 0.0025$&$0.0163 $&$-0.7$&$0.163 $&$-0.7 $\\
$A^{0,\mu}_\mathrm{FB}$&$ $&$0.0169\pm 0.0013$&$ $&$+0.5$&$0.166 $&$+0.3 $\\
$A^{0,\tau}_\mathrm{FB}$&$ $&$0.0188\pm 0.0017$&$ $&$+1.5$&$0.163 $&$+1.5 $\\[0.1cm]\hline
$ $&$ $&$ $&$ $&$ $&$ $&$ $\\[-0.25cm]
$A_e\left(\mbox{SLD}\right)$&$\cite{LEPEWWG06}$&$0.1516\pm 0.0021$&$0.1474$&$+2.0$&$0.1474 $&$+2.0 $\\
$A_\mu\left(\mbox{SLD}\right)$&$ $&$0.142\pm 0.015$&$ $&$-0.4$&$0.1499 $&$-0.5 $\\
$A_\tau\left(\mbox{SLD}\right)$&$ $&$0.136\pm 0.015$&$ $ &$-0.8$&$0.1474 $&$-0.8 $\\[0.1cm]\hline
$ $&$ $&$ $&$ $&$ $&$ $&$ $\\[-0.25cm]
$A_e\left(P_\tau\right)$&$\cite{LEPEWWG06}$&$0.1498\pm 0.0049$&$ $&$+0.5 $&$0.1474 $&$+0.5 $\\
$A_\tau\left(P_\tau\right)$&$ $&$0.1439\pm 0.0043$& &$-0.8$&$0.1474 $&$-0.8 $\\[0.1cm]\hline
$ $&$ $&$ $&$ $&$ $&$ $&$ $\\[-0.25cm]
$R^0_b$&$\cite{LEPEWWG06}$&$0.21629\pm 0.00066$&$0.21581 $&$+0.7 $&$0.21581 $&$+0.7 $\\
$R^0_c$&$ $&$0.1721\pm 0.0030$&$0.1722 $&$0.0 $&$0.1722 $&$0.0 $\\
$A^{0,b}_\mathrm{FB}$&$ $&$0.0992\pm 0.0016$&$0.1033 $&$-2.6 $&$0.1033 $&$-2.6 $\\
$A^{0,c}_\mathrm{FB}$&$ $&$0.0707\pm 0.0035$&$0.0738 $&$-0.9 $&$0.0738 $&$-0.9 $\\ 
$A_b$&$ $&$0.923\pm 0.020$&$0.935 $&$-0.6$&$0.935 $&$-0.6 $\\
$A_c$&$ $&$0.670\pm 0.027$&$0.668 $&$+0.1 $&$0.668 $&$+0.1 $\\[0.1cm]\hline
$ $&$ $&$ $&$ $&$ $&$ $&$ $\\[-0.25cm]
$\sin^2{\theta_\mathrm{eff}^\mathrm{lept}}\left(Q_\mathrm{FB}^\mathrm{had}\right)$&$\cite{LEPEWWG06}$&$0.2324\pm 0.0012$&$0.23148 $&$+0.8 $&$0.23148 $&$+0.8 $\\[0.1cm]\hline
$ $&$ $&$ $&$ $&$ $&$ $&$ $\\[-0.25cm]
$g^2_L$&$\cite{Yao:2006px}$&$0.3005\pm 0.0012$&$0.3038 $&$-2.8 $&$0.3038 $&$-2.8 $\\[0.1cm]
$g^2_R$&$ $&$0.0311\pm 0.0010$&$0.0301 $&$+1.0$&$0.0301 $&$+1.0 $\\
$\theta_L$&$ $&$2.51\pm 0.033$&$2.46 $&$+1.4$&$2.46 $&$+1.4 $\\
$\theta_R$&$ $&$4.59\pm 0.41$&$5.18 $&$-1.4 $&$5.18 $&$-1.4 $\\[0.1cm]\hline
$ $&$ $&$ $&$ $&$ $&$ $&$ $\\[-0.25cm]
$g_V^{\nu e}$&$\cite{Yao:2006px}$&$-0.040\pm 0.015$&$-0.0385  $&$-0.1 $&$-0.0384 $&$-0.1 $\\
$g_A^{\nu e}$&$ $&$-0.507\pm 0.014$&$-0.5065 $&$0.0 $&$-0.5065 $&$0.0 $\\[0.1cm]\hline
$ $&$ $&$ $&$ $&$ $&$ $&$ $\\[-0.25cm]
$Q_W\left(^{133}_{55}\mbox{Cs}\right)$&$\cite{Ginges:2003qt} $&$-72.74\pm 0.46$&$-72.92
$&$+0.4 $&$-72.92 $&$+0.4 $\\[0.1cm]\hline 
\end{tabular}}
\caption{Measurements of the (pseudo) observables included
  in our fit, compared with the best-fit values in the SM and in the
  SM extended by a $\Delta_1$ doublet coupled to the second family.}
\label{Exp-SM}
\end{center}
\end{table}

\begin{table}[p]
\begin{center}
{\scriptsize
\begin{tabular}{|c c||c|c|c|c|c|} \hline
$ $&$ $&$ $&$ $&$ $&$ $&$ $\\[-0.2cm]
Quantity &&Experimental Value& Standard Model&Pull&Extended Model&Pull\\
         &&                  &               &    &with $N$ Universal&\\ \hline \hline
$ $&$ $&$ $&$ $&$ $&$ $&$ $\\[-0.25cm]
$m_t$&$\cite{Tevatronmt}$&$172.6\pm 1.4$&$172.9 $&$-0.2$&$172.7 $&$-0.1 $\\
$\Delta \alpha^{(5)}_\mathrm{had}\left(M_Z^2\right)$&$\cite{Burkhardt:2005se}$&$0.02758\pm 0.00035$&$0.02756 $&$+0.1$&$0.02769 $&$-0.3 $\\[0.1cm]
$\alpha_S\left(M_Z^2\right)$&$\cite{Yao:2006px}$&$0.1176\pm 0.002$&$0.1181 $&$-0.2$&$0.1181 $&$-0.2 $\\[0.1cm]\hline
$ $&$ $&$ $&$ $&$ $&$ $&$ $\\[-0.25cm]
$M_W\left[\mbox{GeV}\right]$&$\cite{Grunewald:2007xt}$&$80.398\pm 0.025$&$80.365 $&$+1.3$&$80.362 $&$+1.4 $\\
$\mbox{Br}\left(W\rightarrow l\nu\right)$&$\cite{Yao:2006px}$&$0.1080\pm0.0009 $&$0.1083 $&$-0.3$&$0.1082 $&$-0.3 $\\[0.1cm]\hline
$ $&$ $&$ $&$ $&$ $&$ $&$ $\\[-0.25cm]
$M_Z\left[\mbox{GeV}\right]$&$\cite{LEPEWWG06}$&$91.1875\pm 0.0021$&$91.1876 $&$-0.1$&$91.1874 $&$0.0 $\\
$\Gamma_Z \left[\mbox{GeV}\right]$&$ $&$2.4952\pm 0.0023$&$2.4952 $&$0.0$&$2.4961 $&$-0.4 $\\
$\sigma_H^0\left[\mbox{nb}\right]$&$ $&$41.540\pm 0.037$&$41.480 $&$+1.6$&$41.501 $&$+1.1 $\\
$R^0_l$&$ $&$20.767\pm 0.025$&$20.738 $&$+1.2$&$20.740 $&$+1.1 $\\
$A^{0,l}_\mathrm{FB}$&$ $&$0.0171\pm 0.0010$&$0.0163 $&$+0.8$&$0.0164 $&$+0.7 $\\[0.1cm]\hline
$ $&$ $&$ $&$ $&$ $&$ $&$ $\\[-0.25cm]

$A_l\left(\mbox{SLD}\right)$&$\cite{LEPEWWG06}$&$0.1513\pm 0.0021$&$0.1474 $&$+1.9$&$0.1479 $&$+1.6 $\\[0.1cm]\hline
$ $&$ $&$ $&$ $&$ $&$ $&$ $\\[-0.25cm]
$A_l\left(P_\tau\right)$&$\cite{LEPEWWG06}$&$0.1465\pm 0.0033$&$ $&$-0.3$&$ $&$-0.4 $\\[0.1cm]\hline
$ $&$ $&$ $&$ $&$ $&$ $&$ $\\[-0.25cm]
$R^0_b$&$\cite{LEPEWWG06}$&$0.21629\pm 0.00066$&$0.21582 $&$+0.7 $&$0.21582 $&$+0.7 $\\
$R^0_c$&$ $&$0.1721\pm 0.0030$&$0.1722$&$0.0 $&$0.1722 $&$0.0 $\\
$A^{0,b}_\mathrm{FB}$&$ $&$0.0992\pm 0.0016$&$0.1033  $&$-2.6 $&$0.1036 $&$-2.8 $\\
$A^{0,c}_\mathrm{FB}$&$ $&$0.0707\pm 0.0035$&$0.0738  $&$-0.9 $&$0.0741 $&$-1.0 $\\ 
$A_b$&$ $&$0.923\pm 0.020$&$0.935  $&$-0.6$&$0.935 $&$-0.6 $\\
$A_c$&$ $&$0.670\pm 0.027$&$0.668  $&$+0.1 $&$0.668 $&$+0.1 $\\[0.1cm]\hline
$ $&$ $&$ $&$ $&$ $&$ $&$ $\\[-0.25cm]
$\sin^2{\theta_\mathrm{eff}^\mathrm{lept}}\left(Q_\mathrm{FB}^\mathrm{had}\right)$&$\cite{LEPEWWG06}$&$0.2324\pm 0.0012$&$0.23148 $&$+0.8 $&$0.23141 $&$+0.8 $\\[0.1cm]\hline
$ $&$ $&$ $&$ $&$ $&$ $&$ $\\[-0.25cm]
$g^2_L$&$\cite{Yao:2006px}$&$0.3005\pm 0.0012$&$0.3038 $&$-2.8 $&$0.3038 $&$-2.8 $\\[0.1cm]
$g^2_R$&$ $&$0.0311\pm 0.0010$&$0.0301 $&$+1.0$&$0.0301 $&$+1.0 $\\
$\theta_L$&$ $&$2.51\pm 0.033$&$2.46 $&$+1.4$&$2.46 $&$+1.4 $\\
$\theta_R$&$ $&$4.59\pm 0.41$&$5.18 $&$-1.4 $&$5.18 $&$-1.4 $\\[0.1cm]\hline
$ $&$ $&$ $&$ $&$ $&$ $&$ $\\[-0.25cm]
$g_V^{\nu e}$&$\cite{Yao:2006px}$&$-0.040\pm 0.015$&$-0.0384 $&$-0.1 $&$-0.0386 $&$-0.1 $\\
$g_A^{\nu e}$&$ $&$-0.507\pm 0.014$&$-0.5065 $&$0.0 $&$-0.5064 $&$0.0 $\\[0.1cm]\hline
$ $&$ $&$ $&$ $&$ $&$ $&$ $\\[-0.25cm]
$Q_W\left(^{133}_{55}\mbox{Cs}\right)$&$\cite{Ginges:2003qt}$&$-72.74\pm 0.46$&$-72.92$&$+0.4 $&$-72.95 $&$+0.5 $\\[0.1cm]\hline 
\end{tabular}}
\caption{Measurements of the (pseudo) observables included
  in our fit assuming lepton universality, compared with the best-fit
  values in the SM and in the SM extended by universal singlets $N$.} 
\label{Exp-SM-LU}
\end{center}
\end{table}
\newpage


\end{document}